\documentclass[reprint,aps,pra,notitlepage,longbibliography,superscriptaddress,floatfix]{revtex4-2}

\usepackage{graphicx}
\graphicspath{{Figures/}}
\usepackage{xr}
\usepackage{amsmath} 
\usepackage{amsfonts}
\usepackage{bbm}
\usepackage{braket}
\usepackage{graphicx}
\usepackage{float}
\usepackage{url}
\usepackage{qcircuit}
\usepackage{amssymb}
\usepackage{qcircuit}
\usepackage{siunitx}
\usepackage{comment}
\usepackage{xcolor}
\usepackage{mwe}    

\makeatletter
\newcommand*{\addFileDependency}[1]{
  \typeout{(#1)}
  \@addtofilelist{#1}
  \IfFileExists{#1}{}{\typeout{No file #1.}}
}
\makeatother

\usepackage{siunitx}
\usepackage[export]{adjustbox}
\usepackage{graphicx}
\usepackage{dcolumn}
\usepackage{bm}
\usepackage{comment}
\usepackage{xr-hyper}
\usepackage{hyperref}

\newcommand{\CNOT}{\mathtt{CNOT}}

\newcommand{\yb}{$^{171}$Yb$^+$ }
\newcommand{\sfz}{$^2$S$_{1/2}{\left|F=0, m_f=0\right>}$ }
\newcommand{\sfo}{$^2$S$_{1/2}{\left|F=1\right>}$ }

\newcommand{\sfoc}{$^2$S$_{1/2}{\left|F=1,m_f=0\right>}$ }

\newcommand{\pfz}{$^2$P$_{1/2}{\left|F=0\right>}$ }

\newcommand{\dukeece}{Department of Electrical and Computer Engineering, Duke University, Durham, NC 27708, USA}
\newcommand{\dukephysics}{Department of Physics, Duke University, Durham, NC 27708, USA}
\newcommand{\dukechemistry}{Department of Chemistry, Duke University, Durham, NC 27708, USA}
\newcommand{\ionq}{IonQ, Inc., College Park, MD 20740, USA}
\newcommand{\dqc}{Duke Quantum Center, Duke University, Durham, NC 27701, USA}

\begin{document}

\title{Hidden Inverses: Coherent Error Cancellation at the Circuit Level}

\author{Bichen Zhang}
\email{bichen.zhang@duke.edu, ken.brown@duke.edu}

\author{Swarnadeep Majumder}
\affiliation{\dqc}
\affiliation{\dukeece}

\author{Pak Hong Leung}
\affiliation{\dqc}
\affiliation{\dukephysics}

\author{Stephen Crain}
\altaffiliation{Present Address: \ionq}

\author{Ye Wang}

\author{Chao Fang}
\affiliation{\dqc}
\affiliation{\dukeece}

\author{Dripto M. Debroy}
\altaffiliation{Present Address: Google Research, Venice, CA 90291, USA}
\affiliation{\dqc}
\affiliation{\dukephysics}

\author{Jungsang Kim}
\affiliation{\dqc}
\affiliation{\dukeece}
\affiliation{\dukephysics}
\affiliation{\ionq}

\author{Kenneth R. Brown}
\email{bichen.zhang@duke.edu, ken.brown@duke.edu}
\affiliation{\dqc}
\affiliation{\dukeece}
\affiliation{\dukephysics}
\affiliation{\dukechemistry}

\begin{abstract}
Coherent gate errors are a concern in many proposed quantum computing architectures. Here, we show that certain coherent errors can be reduced by a local optimization that chooses between two forms of the same Hermitian and unitary quantum gate. We refer to this method as \textit{hidden inverses}, and it relies on constructing the same gate from either one sequence of physical operations or the inverted sequence of inverted operations. We use parity-controlled $Z$ rotations as our model circuit and numerically show the utility of hidden inverses as a function of circuit width $n$. We experimentally demonstrate the effectiveness for $n=2$ and $n=4$ qubits in a trapped-ion quantum computer. We numerically compare the method to other gate-level compilations for reducing coherent errors.
\end{abstract}

\maketitle

\section{INTRODUCTION}
\label{sec:intro}
Coherent errors are common in quantum computers due to imperfect classical control and parameter drift during the execution of quantum algorithms. Coherent errors can be suppressed by dynamic decoupling \cite{hahn1950, viola1999, biercuk2009optimized, qi2017optimal, quiroz2013, lidar2014review, zeng2018general}, composite pulses \cite{merrill2014progress,low2014,low2016, MountPRA2015}, dynamically corrected gates \cite{KhodjastehPRL2009,ButerakosPRXQ2021}, or randomized compiling \cite{wallman2016}, as well as through decoherence-free subspaces \cite{lidar1998, kwiat2000, lidar2014review, debroy2021optimizing} and quantum error correction \cite{debroy2018, beale2018, huang2019performance, iverson2020}. When we have more information about the structure exhibited by the coherent errors, we can more efficiently suppress them. For many systems, accurate classical control and calibration of the driving fields used to manipulate qubits remain a critical challenge~\cite{Bruzewicz2019apr,kellyarXiv2018,BaumPRXQuantum2021}. Fluctuating environmental conditions can lead to slow drifts of system parameters over the course of many experimental runs.

Here we present a method for canceling coherent errors without requiring any additional gates. We call this method {\em hidden inverses} because it uses the fact that quantum circuits often contain inverted gates, which are hidden due to the common use of self-adjoint unitary operations, for example, the Hadamard ($H$), controlled-NOT ($\mathtt{CNOT}$), and Toffoli gates. Since these gates are constructed from physical operations, it can be beneficial to choose between different configurations. As an example, we could choose from two $\CNOT$ sequences, $\mathtt{CNOT}_a=ABC$ or $\mathtt{CNOT}_b=\mathtt{CNOT}_a^\dagger=C^\dagger B^\dagger A^\dagger$, where A, B, C, $A^\dagger$, $B^\dagger$ and $C^\dagger$ are native physical gates. The goal is to determine when to implement $\mathtt{CNOT}_a$, and when to implement $\mathtt{CNOT}_b=\mathtt{CNOT}_a^\dagger$. This decision will depend on both the neighboring gates and the underlying error process.


The paper is organized as follows: In  Sec. \ref{sec:theory} we introduce the idea of hidden inverses and discuss their application in a common quantum circuit structure, the parity-controlled $Z$ rotation widely used to create high-weight Pauli operations \cite{Nielsen2020,whitfield2011simulation}. In Sec. \ref{sec:exp_system}, we give an overview of the trapped ion quantum computing experimental platform used for the demonstration of the hidden inverse technique. In Sec. \ref{sec:exp_result}, we present the experimental results and analyze the performance of hidden inverses.  In Sec. \ref{sec:altmethods}, we numerically compare hidden inverses to other methods for reducing coherent errors. Finally, in Sec. \ref{sec:conclusions}, we summarize the results and conclude.

\section{Designing gates to control systematic error }
\label{sec:theory}

For solid-state and atomic quantum systems, quantum gates are performed by applying electromagnetic signals that generate a time-dependent Hamiltonian. For small systems, the advantage of a gate description is questionable and direct optimization of the pulse sequences to optimize the overall evolution is preferable \cite{KhanejaPhysRevA2001, ShiASPLOS2019, MuraliASPLOS2019, GokhaleMicro2019, MagannPRXQuantum2021, meiteiArXiv2021}. Gates become a useful abstraction when we consider both larger systems and the challenge of calibration for multiple applications. 

Let $U_0$ be the ideal gate and $U_a$ be an instance of the gate. $U_a$ could be a completely positive trace-preserving map, but we invoke the Stinespring dilation theorem to say that all instances of the gate differ only from the ideal gate by a unitary operation that is potentially on a larger Hilbert space.  We consider both the left and right error operators $V$, $U_a=U_0V_{R,a}=V_{L,a}U_0$. We are interested in the case where $U_a$ is a good approximation of $U_0$ and therefore $V_{R}$ and $V_{L}$ must be close to the identity.

The system controller does not control the noise but chooses only what signals to imperfectly apply. We further break the unitary label into a control label and an error label $U_{a,\epsilon}$. The question that we are exploring is whether it makes sense for the controller to have multiple versions of $U$, or only a single version, and given multiple versions what are the methods for compilation.

A very common motif in quantum mechanics and quantum algorithms is a unitary $W$ that is transformed by another unitary $U$, $\tilde{W}=UWU^\dagger$. We should find methods for implementing $U_a$ and $U_b^\dagger$  such that when averaged over the noise instances, $\epsilon$ and $\delta$,  $U_{a,\epsilon}WU_{b,\delta}^\dagger$ is as close to $\tilde{W}$.  This will work perfectly, if $V_{R,a,\epsilon} W V^\dagger_{L,b,\delta}=W$.  The ideal cases are either when there is no noise or  when $W$ commutes with the $V$'s and $V^\dagger_{R,a,\epsilon}=V^\dagger_{L,b,\delta}$.  In many cases $W$ itself is a small rotation, for example, in a Trotter series,  and  $V^\dagger_{R,a,\epsilon}=V^\dagger_{L,b,\delta}$ will still yield a reduction in error.

Given a quantum circuit with the structure $\tilde{W}=UWU^\dagger$ one can choose the appropriate $U_a$ and $U_b^\dagger$ for every $W$. We limit ourselves in this study to the case of ``hidden inverses" where $U=U^\dagger$, which occurs regularly in quantum circuits, for example,  when operations are conjugated by $\CNOT$s. Our key observation is that in many physical systems, a $\CNOT$ is constructed from a series of system-specific gates, whose inverses are readily available by changing the sign of the control field \cite{maslov2017, MadzikNatComm2021}. In these cases, $\CNOT_a$ corresponds to the regular order of gates with the regular control fields, and $\CNOT_b$ inverts the sequence and inverts the control field. It is the driven Hermitian conjugate of $\CNOT_a$. This extends to any unitary, which is its self-inverse.

\subsection{Parity-controlled rotations and theoretical error models}
A common place where self-inverses arise is when multiple $\mathtt{CNOT}(c,t)$s are used to conjugate a single-qubit rotation to generate a multiqubit unitary. This structure is common in quantum simulation and quantum optimization algorithms \cite{whitfield2011simulation, barkoutsos2018}.  To generate an $n$ qubit weight Pauli operation, we can build a circuit that performs a single-qubit $Z$, $Z(\theta)=\exp\left[-i\frac{\theta}{2}Z\right]$, rotation whose direction is conditioned on the parity of $n-1$ other qubits as follows:
\small{
\begin{eqnarray}
    \nonumber
    U &=& \left[\prod_{j=1}^{n-1} {\CNOT}(j,n)\right] Z_{n}(\theta)\left[ \prod_{k=1}^{n-1} {\CNOT}(n-k,n)\right]\\
    &=&\exp(-i\frac{\theta}{2}\bigotimes_{j=1}^{n}Z_j)
    \label{eq:multi_qubit_cz}.
\end{eqnarray}
}

\begin{figure}
\includegraphics[width=\columnwidth]{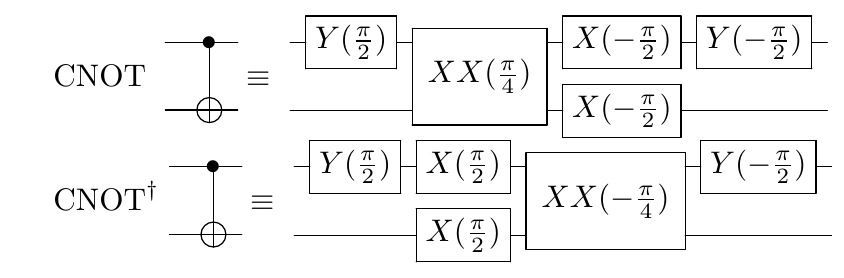}
	\caption[]{Standard and Hermitian conjugated decompositions of $\CNOT$ gate with native trapped-ion quantum operations consisting of single-qubit gates
	and M\o lmer-S\o rensen interactions. }
		\label{fig:cnot_decomp}
\end{figure}

We expect the $\CNOT$s that come before the rotation and the $\CNOT$s that come after the rotation should differ depending on the error model.  At this point, we need to introduce a physical decomposition of the $\CNOT$s and an error model to proceed. Many error models can be considered. In the main text, we limit ourselves to a model where the entangling operation is generated by a single two-qubit Pauli operator that is decorated with single-qubit gates, but we also consider the gate Hamiltonian model introduced in Ref. \cite{BarnesPRA2017} in Appendix \ref{app:toymodel}. For concreteness we further specialize to an $XX$-type interaction common in M\o lmer-S\o rensen (MS) gates in trapped ions \cite{sorensen1999}. For this case, the $\CNOT$ gate can be performed in either its standard or hidden inverse configurations as displayed in Fig. \ref{fig:cnot_decomp} with ion trap gates $XX(\theta)=\exp(-i\theta XX)$, $X(\theta)=\exp[-i\frac{\theta}{2}X]$, and $Y(\theta)=\exp[-i\frac{\theta}{2}Y]$. The choice of configuration is nontrivial since $\CNOT$ gates are synthesized from a two-qubit gate and multiple single-qubit gates \cite{maslov2017}, which are subject to systematic overrotations and phase errors in addition to stochastic noise.

We first consider a simplified error model where only the two-qubit M\o lmer-S\o rensen gates have the same overrotation by a fraction $\epsilon$. We calculate the average gate fidelity, $F$, from the entanglement fidelity of two unitaries $U$ and $V$, $F_e=|\mathrm{Tr}[U^\dagger V]|^2/4^n$, for the $(n-1)$-qubit parity-controlled rotation (Eq. \ref{eq:multi_qubit_cz}) as a function of $\theta$, $\epsilon$, and $n$ to be 
\begin{eqnarray}
F_{e,\pm}(\theta,\epsilon,n)&=&\left[\cos\left(\frac{\pi}{4}\epsilon\right)^2\pm \sin\left(\frac{\pi}{4}\epsilon\right)^2\cos\left(\theta\right)\right]^{2(n-1)}\\
F_{\pm}(\theta,\epsilon,n)&=&\frac{2^{n}F_{e,\pm}(\theta,\epsilon,n)+1}{2^{n}+1},
\end{eqnarray}
where $F_+$ is the hidden inverse configuration fidelity and $F_-$ is the standard configuration fidelity. We note that $F_+(\theta,\epsilon,n)=F_-(\theta+\pi,\epsilon,n)$ and we can obtain the best fidelity by choosing hidden inverses for $|\theta|\leq\pi/2$ and the standard circuit for  $|\theta|>\pi/2$ for $\theta\in [-\pi,\pi]$. At $\theta=0$ and $\theta=\pi$, there is an exponential difference in the fidelity between the two choices as a function of $n$. For small $\epsilon$ and small angle deviation, $\vartheta$ , around these ideal points of 0 and $\pi$, for the correct sequence choice fidelity drops as $(n-1)\left(\pi/4\right)^2\epsilon^2\vartheta^2$, while for the incorrect sequence the fidelity drops as $(n-1)\left({\pi}/{4}\right)^2\epsilon^2\left(4-\vartheta^2\right)$. We only achieve perfect cancellation at the ideal points, but we benefit from our choice as long as $\theta$ is close to the ideal point. 

With this simplified error model in mind, we consider additional errors starting with overrotation errors on all gates, phase-misalignment errors, and then a consideration of these systematic errors with additional stochastic errors. Phase misalignment occurs because two-qubit and single-qubit gates are driven by different fields and mechanisms. The $Z$ basis is well defined by the energy eigenbasis of the undriven system Hamiltonian. $X$ and $Y$ in the rotating frame simply differ by a phase and a common experimental challenge is to align the $X$ in a two-qubit $XX$ interaction with the single qubit $X$ interaction.

\begin{figure}[tp]
	\begin{center}
		\includegraphics[width=\columnwidth]{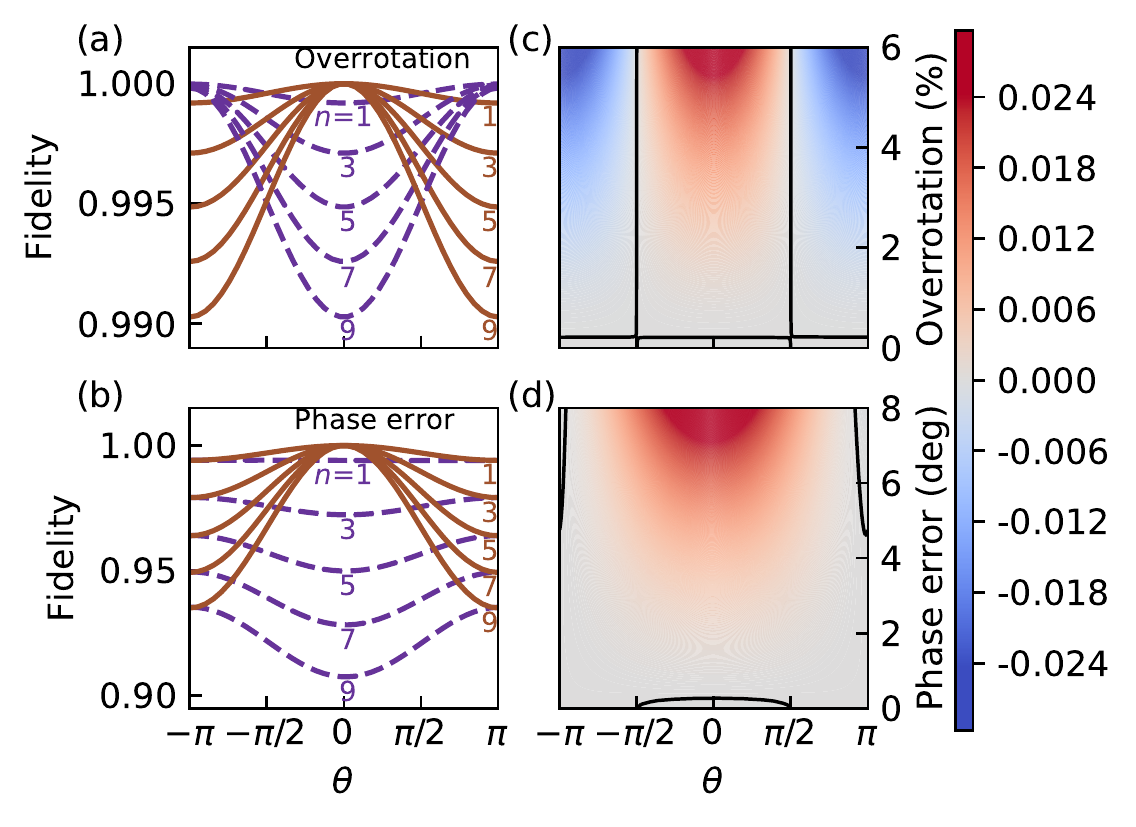}
		\caption{ (a) Average gate fidelities of parity controlled-$Z$ rotation circuits according to Eq. \ref{eq:multi_qubit_cz} with standard (dashed lines) and hidden inverse (solid lines) configurations for circuit width $n \in \{2,4,...,10\}$, with each two-qubit gate subject to $\epsilon_{2Q}=2\%$ overrotation and each single-qubit gate subject to $\epsilon_{1Q}=0.2\%$ overrotation. The hidden inverse configuration outperforms the standard one when the absolute value of the $Z$ rotation angle is less than approximately $\pi/2$.  (b) Average gate fidelities of parity controlled-$Z$ rotation circuits with $\phi_{\rm{diff}}=3.5^\circ$ phase misalignment. The performance of hidden inverse configuration is higher than or equal to the standard one regardless of $\theta$. This shows the scalability of the hidden inverse technique. (c) (d) The average gate-fidelity difference between hidden inverse and standard configuration when circuit width $n=2$. The warm (cool) color shows the area where hidden inverse configuration outperforms (underperforms) the standard one. Curves in black represent the boundary where the fidelity difference is zero. When the coherent error is small, fidelities of standard configuration surpass the hidden inverse configuration by approximately $10^{-5}$ due to the imperfect motional state coherence. }
		\label{fig:Simulation}
	\end{center}
\end{figure}
In Fig. \ref{fig:Simulation}, we examine how these errors affect the circuit fidelity of implementing Eq. \ref{eq:multi_qubit_cz} using the standard circuit with only $\CNOT$ and the hidden inverse circuit using $\CNOT$ and $\CNOT^\dagger$ from Fig. \ref{fig:cnot_decomp}.  These choices impact circuit performance when each gate is subject to either only overrotation error (Fig. \ref{fig:Simulation}(a)) or only a phase misalignment between single- and two-qubit gates (Fig. \ref{fig:Simulation}(b)).

In both Fig. \ref{fig:Simulation}(a) and (b), we observe that the hidden inverse configuration outperforms the standard one considerably when the $Z(\theta)$ rotation angle is small. This can be understood by noting a small angle rotation is near the identity, so the systematic errors are approximately canceled. We further note the oscillatory behavior suggests the need for compilation tools to determine which $\CNOT$s should be inverted in more complex circuits. The amplitude of the oscillation is positively correlated with the number of control qubits showing the affect of these choices increases with circuit size. Due to the ubiquity of $\CNOT$ conjugations about single-qubit rotations in quantum algorithms, we expect a peep-hole style optimization \cite{McKeeman65} would work well when slowly drifting overrotation or phase offset errors dominate. 

In any real system, there will be multiple error types and the potential advantage can differ. Here we now examine numerically the $n=2$ case using a detailed ion-trap error model. For the simulations shown in Fig. \ref{fig:Simulation}(c) and (d), we consider the systematic overrotation error, phase misalignment, and all dominant stochastic error sources in our experimental system \cite{wang2020} including laser dephasing error, motional dephasing error, and motional heating. We use a master equation in Lindblad form to simulate the open quantum system. The dominant stochastic error sources are depicted by corresponding collapse operators, while the coherent error sources are represented by parameter offsets in the Hamiltonian. Further details can be found in Appendix \ref{sec:errormodel}. 

For zero phase-alignment error, we compare the relative fidelity difference between standard circuits and hidden inverse circuits as we change the overrotation angle in Fig. \ref{fig:Simulation} (c).  We see that the broad feature of the overrotation data in Fig. \ref{fig:Simulation}(a) is preserved and hidden inverses perform well for $|\theta|<\pi/2$.  For small systematic errors, the difference in fidelity between the two circuits and the ideal circuit is less than $10^{-5}$, while the circuit fidelities are close to $1-10^{-2}$.

Fig. \ref{fig:Simulation}(d) shows that the hidden inverse configuration suppresses phase misalignment in almost all the area of interest, even with additional stochastic noise. In the small region where the hidden inverse configuration exaggerates the phase misalignment, the fidelity difference is at the level of $10^{-5}$. Unlike for overrotation errors, where the advantage of the hidden inverse sequence requires the $Z(\theta)$ rotation angle to be small, the hidden inverse configuration provides a fidelity improvement for most phase-misalignment errors given our gate model.

\begin{figure*}[htp]
\includegraphics[width=\textwidth]{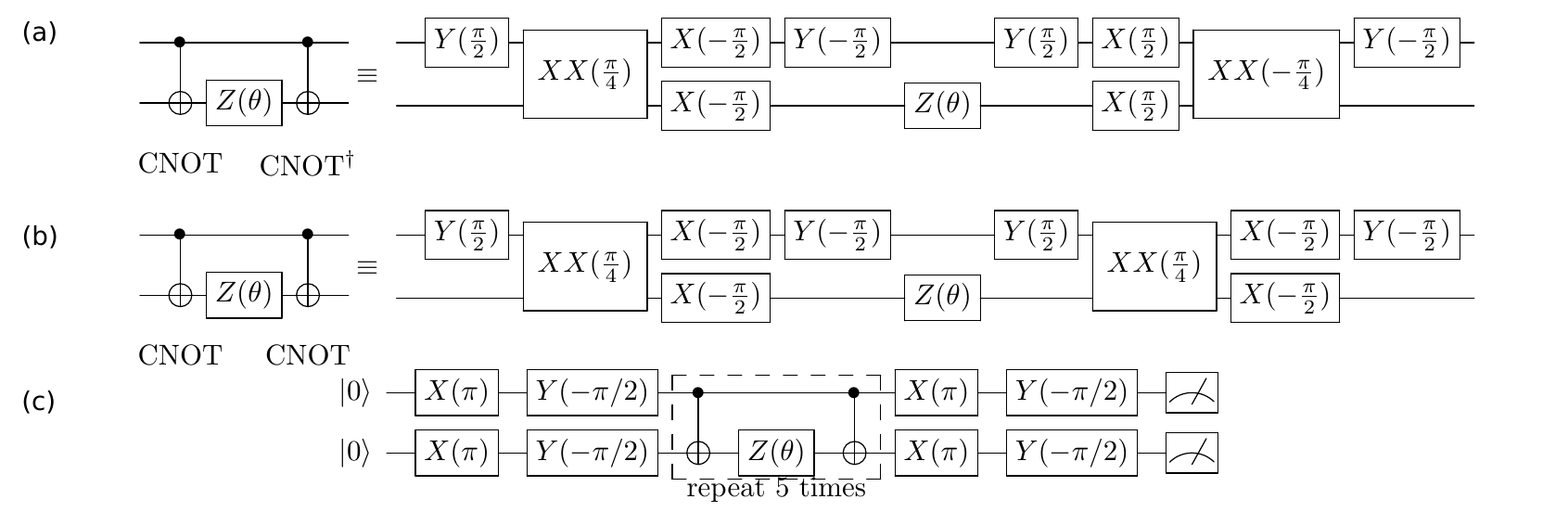}
	\caption{ (a) Decomposition of the hidden inverse parity-controlled $Z$ rotation circuit ($n=1$)  into native gates for trapped-ion qubits where the second $\CNOT$ gate has a 180$^\circ$ relative phase shift with respect to the first $\CNOT$ gate. (b) Decomposition of the standard circuit into trapped-ion qubit operations where the second $\CNOT$ gate is the same as the first one. (c) The two-qubit experimental circuit for investigating the impact of hidden inverses. The portion of the circuit highlighted by the dashed box is repeated 5 times in order to amplify the effects of the coherent errors.}
		\label{fig:cnot_circuit_diagrams}
\end{figure*}

\section{Experiment implementation of an arbitrary quantum circuit}
\label{sec:exp_system}

In the experiment, a chain of \yb ions is trapped in a linear chain \SI{70}{\um} above the surface of a microfabricated surface trap made by Sandia National Laboratories. The $\ket{0}$ and $\ket{1}$ states of the qubit are encoded in the hyperfine ground states, \sfz and \sfoc, respectively. A 369.5-nm laser is used to Doppler cool, electromagnetically-induced-transparency (EIT) cool, and prepare the ions in the $\ket{0}$ state. State detection is performed through state-dependent fluorescence by resonantly exciting the \sfo to \pfz transition and collecting the emitted photons~\cite{Noek2013,Myerson2008}. The scattered photons are imaged with a 0.6 numerical-aperture lens and coupled into a linear array of multimode fibers with 100-$\mu$m-diameter cores~\cite{crain2019high}. Each fiber in the array is connected to individual photomultiplier tubes, allowing for individual qubit readout. For the following experiments the Doppler cooling, EIT cooling, state initialization, and state detection take \SI{1}{ms}, \SI{500}{\micro s}, \SI{15}{\micro s}, and \SI{300}{\micro s}, respectively.
Stimulated Raman transitions using a 355-nm picosecond pulsed laser drive single-qubit and two-qubit gates~\cite{wineland1998experimental,Mount2013,inlek2014}. An elliptical beam addresses all qubits in the chain simultaneously while two tightly focused beams perpendicular to the elliptical beam individually address the two qubits~\cite{crain2014individual}. Steering of each individual beam over the ion chain is accomplished by a pair of micro-electromechanical systems (MEMS) mirrors each tilting in orthogonal directions. The number of atomic qubits in our trapped-ion quantum processor is limited to $13$ by the steering range of the MEMS mirrors. In this paper, a two-ion chain (two-qubit circuits) and a five-ion chain (four-qubit circuits) are used to prove the principle. The beams pass through acousto-optic modulators (AOMs) driven by a radio frequency system on chip (RFSoC), which provides the ability to change the amplitude, frequency, and phase of each beam. The RFSoC firmware is provided by Sandia National Laboratories QSCOUT project \cite{Clark2021}. By controlling the duration of the pulse and the phase of one of the two Raman beams we can perform arbitrary single-qubit rotations, $R(\theta , \phi )$. Two-qubit gates are implemented using the M\o lmer-S\o rensen scheme~\cite{Sorensen2000}. Frequency modulation (FM) of the Raman beams is performed in order to robustly disentangle the the qubit states from all of the motional modes~\cite{Leung2018, Landsman2019, wang2020, kang2021batch}. Further details of the setup can be found in Ref \cite{wang2020}.

One universal gate set of our system contains the M\o lmer-S\o rensen [$XX(\pi/4)$] gate, $X(\theta)$ gates, and arbitrary $Z$ rotations. Arbitrary $Z(\theta)$ rotations are implemented in a virtual way by accumulating a $-\theta$ phase in the subsequent gate operations. Indeed, $Y(\theta)$ gates are simply phase-shifted $X(\theta)$ gates.

We implement the M\o lmer-S\o rensen gate in a spin-phase-sensitive configuration \cite{inlek2014}. In this configuration, the rotation axis is not exactly aligned with the $XX$ axis due to mechanical fluctuation in the optical path of the Raman beams. We note the intrinsic phase instability and the slight difference of the ac Stark shift between single- and two-qubit gates, and it is necessary to calibrate the phase between these gates. The calibration is done using parity measurement: We initialize the qubits in $\ket{00}$ state and implement a $XX(\pi/4)$ gate on them. Then we apply a single-qubit $\pi/2$ rotation on both qubits. The phase $\phi$ of the single-qubit gates is varied from $0$ to $2\pi$. Finally, both qubits are measured in $Z$ basis. The measured parity $\mathcal{P}$ is fitted to a sinusoid, $\mathcal{P} = A \cos{(\phi_0+2\phi)}$, with a phase offset $\phi_0$ from the parity-measurement results. In experiments, we observe the phase offset drifts as much as $4^\circ$ in a two-qubit system within several hours. Right after calibration, we can reduce the misalignment to as low as approximately $0.2^\circ$.

We know due to the limits of our ability to stabilize laser intensity and phase at the ion that there will be systematic errors between gates. For single-qubit gates, we use gate set tomography (GST) on both direct quantum pulses and composite quantum pulses to  characterize systematic errors. The results of GST infer time-varying overrotations exist in our system that are stable {for a time $>1$ ms}. Details of the single-qubit GST experiment can be found in Appendix \ref{sec:gst}.


\section{Hidden inverse experimental performance}
\label{sec:exp_result}
\begin{figure*}[htp]
	\begin{center}
		\includegraphics[width=\textwidth]{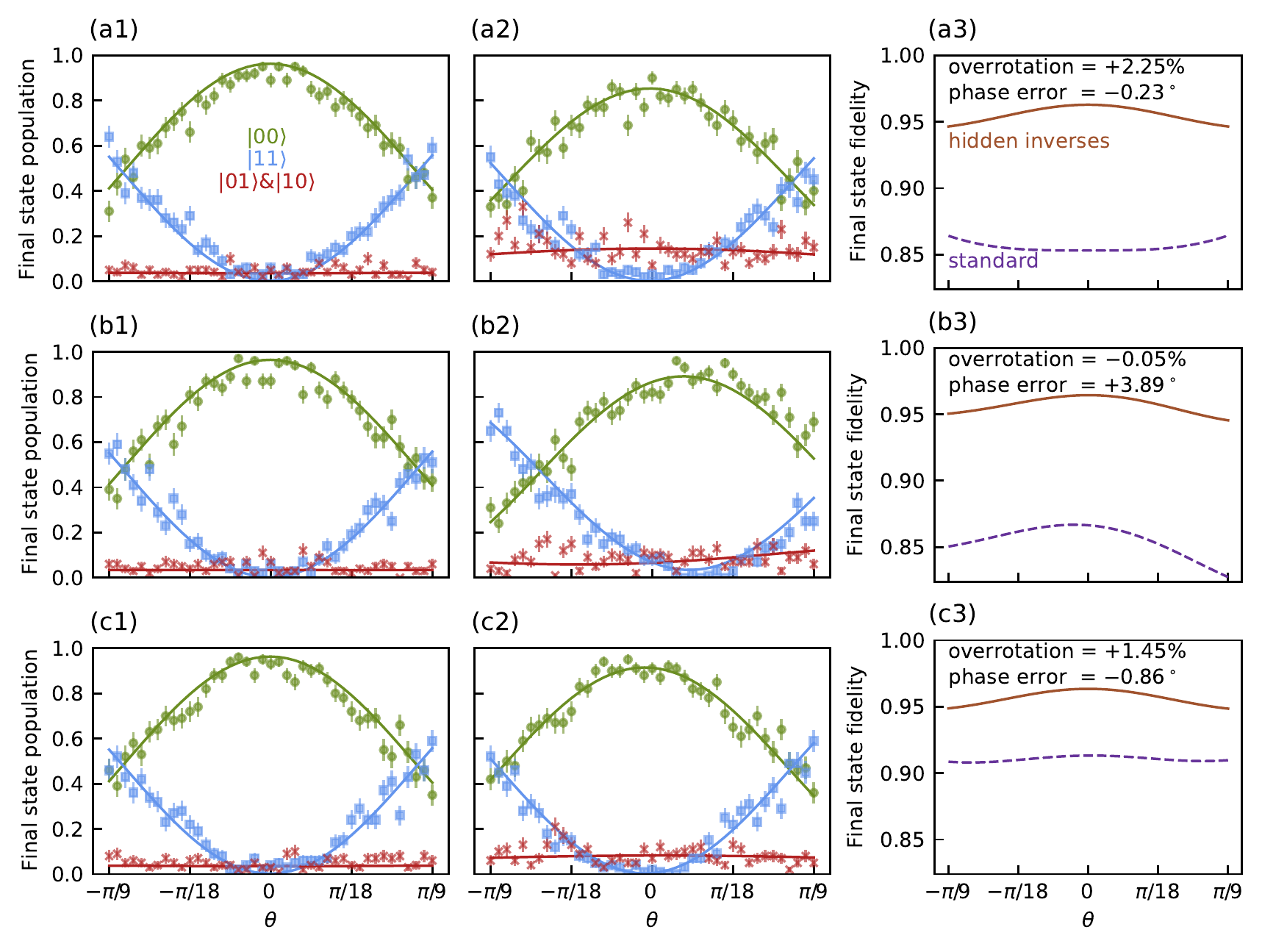}
		\caption[]{Two-qubit hidden inverse experiment data and simulated final state fidelity. The top row (a), the $XX(\pi/4)$ gate experiences {$\epsilon_{2Q}=2.25\pm0.04\%$} overrotation error. The middle row (b), a {$\phi_{\rm{diff}}=3.89 \pm 0.09^\circ$} phase misalignment is introduced to the circuits instead.  Column (1) presents results for the case where our hidden inverse method is implemented, inverting the second $\CNOT$, while column (2) presents results for the standard case where the $\CNOT$ is left unchanged. The green points represent the population of the $\ket{00}$ state, the red points represent the $\ket{01}$ and $\ket{10}$ state, and the blue points represent the $\ket{11}$ state. Solid lines show the results of the simulation with the two types of coherent errors, as well as other dephasing errors. The error bars represent the standard error of the data. Column (3) shows the simulated final state fidelity of the circuits with and without hidden inverses. Hidden inverse circuits outperform the original circuits in both conditions. {The bottom row (c) shows the most ``ideal" results after a full calibration process, during which both types of control error are suppressed as much as possible. An $\epsilon_{2Q}=1.45\pm0.06\%$ overrotation error and a $\phi_{\rm{diff}}=-0.86 \pm 0.14^\circ$ phase misalignment are observed. The $\epsilon_{2Q}=1.45\pm0.06\%$ overrotation error causes a 0.05\% average gate-fidelity drop for one two-qubit gate.}}
		\label{fig:hidden_inverse_data}
	\end{center}
\end{figure*}

The base circuit for the hidden inverse experiment is the portion highlighted by the dashed box in Fig. \ref{fig:cnot_circuit_diagrams}(c). A $\CNOT$ gate is performed followed by a $Z(\theta)$ rotation on the target qubit~\cite{wang2020, maslov2017}. The second $\CNOT$ gate is applied either with the same phase as the first ($\CNOT$) or with a phase shift of $\pi$ relative to the first ($\CNOT^{\dagger}$). The latter configuration is the hidden inverse case. We reverse the gate sequence order and each element gate's sign in the $\CNOT$ decomposition to conform to the Hermitian adjoint's antidistributive  property. The base circuit is repeated 5 times to amplify the two-qubit gate overrotation error and phase-misalignment error between single-qubit and two-qubit gates, which are the dominant coherent error sources in the circuit. We note for the repeated circuits, we cannot experimentally distinguish cancellation of $\CNOT$ and $\CNOT^\dagger$ errors across $Z(\theta)$ with cancellation from the next $\CNOT$. However, the circuit in Fig. \ref{fig:cnot_circuit_diagrams}(c) is needed to amplify the error. 

Two separate sets of experiments are conducted to characterize the two-qubit gate overrotations, phase misalignment between single-qubit and two-qubit gate, and the effectiveness of the hidden inverse scheme. In both sets of experiments, the two-qubit gate fidelity is approximately $99.4\%$ before injecting the coherent errors. The $Z(\theta)$ rotation angle is varied. For the first set of experiments, we introduce a {$\epsilon_{2Q}=2.25\pm0.04\%$} two-qubit gate overrotation error into the circuits and maintain the phase misalignment as small as possible {($\phi_{\rm diff}=-0.23\pm0.17^{\circ}$)}. The circuit in Fig. \ref{fig:cnot_circuit_diagrams}(c) is implemented with and without the hidden inverses to quantify the suppression of overrotation errors. The system can be seen to significantly suffer from two-qubit overrotation error. Fig. \ref{fig:hidden_inverse_data}(a1) shows the probability of detecting the $\ket{00}$, $\ket{10}\&\ket{01}$, and $\ket{11}$ states at the end of the circuit with the hidden inverses, and Fig. \ref{fig:hidden_inverse_data}(a2) shows the results without the hidden inverses. The solid lines indicate fitted simulation results with two free variables, the overrotation error of the $XX$ gates and the phase misalignment between single-qubit and two-qubit gates. When hidden inverses are used, the contrast of the $\ket{00}$ is improved, and the residual population in the odd-parity states is significantly reduced. This indicates suppression of overrotation errors from the $XX$ gates. Using the theoretical model and fitting results, final state fidelities of the circuits in both configurations are estimated. As shown in Fig. \ref{fig:hidden_inverse_data}(a3), the final state fidelities are improved from approximately $85\%$ to $95\%$ due to the usage of hidden inverses. While the $Z(\theta)$ rotation angle increases, the improvement results from hidden inverses decreases.

In the second set of experiments, we introduce a phase-misalignment error of {$\phi_{\rm diff}=3.89 \pm 0.09^{\circ}$} and minimize the two-qubit gate overrotation {($\epsilon_{2Q}=-0.05 \pm 0.22\%$)}. We implement the circuit in both configurations to examine the suppression of phase-misalignment errors for the hidden inverse circuit. Fig. \ref{fig:hidden_inverse_data}(b1) and (b2) show the results of the circuits with and without hidden inverses, respectively, for the set of experiments when phase misalignment is dominant. With the hidden inverse configuration, along with the improved contrast of the $\ket{00}$ population and the reduced odd-parity population, the curves regain symmetry about the $0^{\circ}$ $Z$ rotation. This shows a correction of the phase misalignment between single-qubit gates and two-qubit gates. Fig. \ref{fig:hidden_inverse_data}(b3) represents the estimated final state fidelities of the circuits in both configurations. The fidelities are improved from approximately $84\%$ to $95\%$. In the case of phase misalignment, we note the improvement from hidden inverses fades away much slower than the case of overrotation error as the $Z(\theta)$ rotation angle increases. It agrees with the analysis in Sec. \ref{sec:theory}.

Limited by the systematic error drifts in the experiment system and finite calibration time, we are not able to suppress all coherent error to optimal at the same time. A trade-off between amplitude error and phase error exists. After the most ``ideal" calibration, we observe overrotation $\epsilon_{2Q}=1.45(6) \%$ and phase misalignment $\phi_{\rm{diff}}=-0.9(1)^{\circ}$. Data presented in Fig \ref{fig:hidden_inverse_data}(c) shows that a clear fidelity improvement from hidden inverse configuration is observed in the most ``ideal" condition of our system. We note that the fitting for all two-qubit circuit results is done utilizing the error model described in Appendix \ref{sec:errormodel}.

\begin{figure}
    \centering
    \includegraphics[width=\columnwidth]{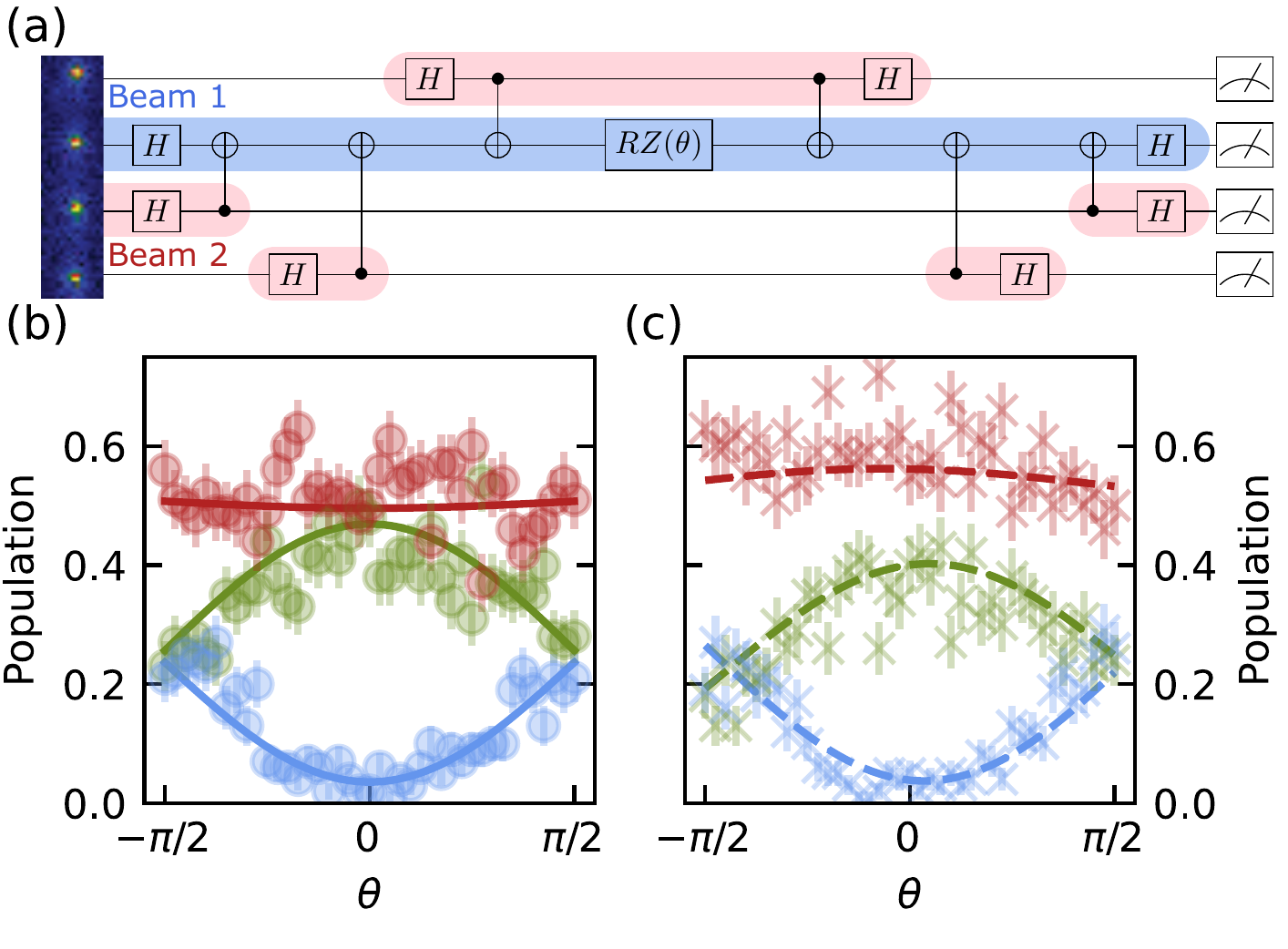}
    \caption{(a) An $n=4$ experimental quantum circuit. The initial Hadamards transform the qubits from $\ket{+}^{\otimes 4}$ to $\ket{+}^{\otimes 4}$, the $\CNOT$s and $Z(\theta)$ implemement Eq. \ref{eq:multi_qubit_cz}, and the final Hadamards map the $X$ basis to the $Z$ basis for measurement. The blue and the pink highlights show the individual qubit addressing scheme by steering tight-focused addressing beam. \cite{wang2020} (b) Probabilities of measuring the four-qubit system in $\ket{0000}$ (green dots), $\ket{1111}$ (blue dots), and others states (red dots) with hidden inverse configuration. The error bars represent the standard error of the data. The data is fitted with a model consists of coherent errors and depolarizing channel. (c) State readout probabilities for standard configuration. The color map is the same as (b).}
    \label{fig:result_5q}
\end{figure}

Lastly, we extend the multiqubit parity control $Z$ circuit to width $n=4$, which is illustrated in Fig. \ref{fig:result_5q}(a). We note the $n=4$ experiments are done in a five-ion chain, with one edge ion qubit idling during the experiment.  With two individual addressing beams, we access the two additional ion qubits by steering one addressing beam with MEMS mirrors \cite{wang2020}. $XX$ gates for all three ion pairs are calibrated separately. The average $\CNOT$ gate fidelity is approximately $90\%$. We assign this fidelity deduction to increasing optical crosstalk ($> 3\%$), optical power loss from MEMS mirrors at large steering angles, and other error sources to be investigated. Similarly, varying the rotation angle $\theta$, we measure final state probabilities for all 16 computational basis states. Fig. \ref{fig:result_5q}(b) and (c) present the final state results utilizing hidden inverses and standard configuration, respectively. By suppressing overrotation error, the hidden inverse configuration improves the contrast of $\ket{0}^{\otimes 4}$ from approximately $0.40$ to approximately $0.47$ and suppresses the average residual population of  states other than $\ket{0}^{\otimes4}$ and $\ket{1}^{\otimes4}$ from approximately $0.55$ to approximately $0.50$. Moreover, hidden inverses help the data points regain symmetry about $\theta=0$, which indicates a correction against phase misalignment. We fit the four-qubit circuit results with a model consisting of coherent error (parameter offsets) and stochastic error (depolarizing channels). The qubits go through a depolarizing channel after every two-qubit gate: with probability $p=0.87$, the state remains the same, while with probability $1-p=0.13$, the state collapses to a totally mixed state. From the fitting where we assume all two-qubit gates experience the same noise channel, we estimate the overrotation $\epsilon_{2Q}\approx5\%$ and the phase misalignment $\phi_{\rm {diff}}\approx -8^\circ$. Due to the strong stochastic noise from inter-beam crosstalk \footnote{Unlike the coherent crosstalk noise caused by a single addressing beam, the finite phase coherence time between two addressing beams makes the interbeam crosstalk stochastic noise after averaging over multiple experiment shots.}, hidden inverses provide only a limited improvement but still at no experimental cost.



\section{Alternative methods for reducing systematic errors}
\label{sec:altmethods}
Systematic errors can be reduced in a number of ways, and we briefly compare our method with other techniques in the context of the experiment. Hidden inverses work well in experiments with multiple $\CNOT$s and with compatible systematic errors, even with some stochastic noise.  It cannot be as powerful as total circuit optimization, but it provides a local control solution that can be applied to any quantum computer without additional time overhead.

\subsection{Two-qubit Solovay-Kitaev-1 (SK1) composite pulses}
\begin{figure}[tp]
	\begin{center}
 	    \includegraphics[width=\columnwidth]{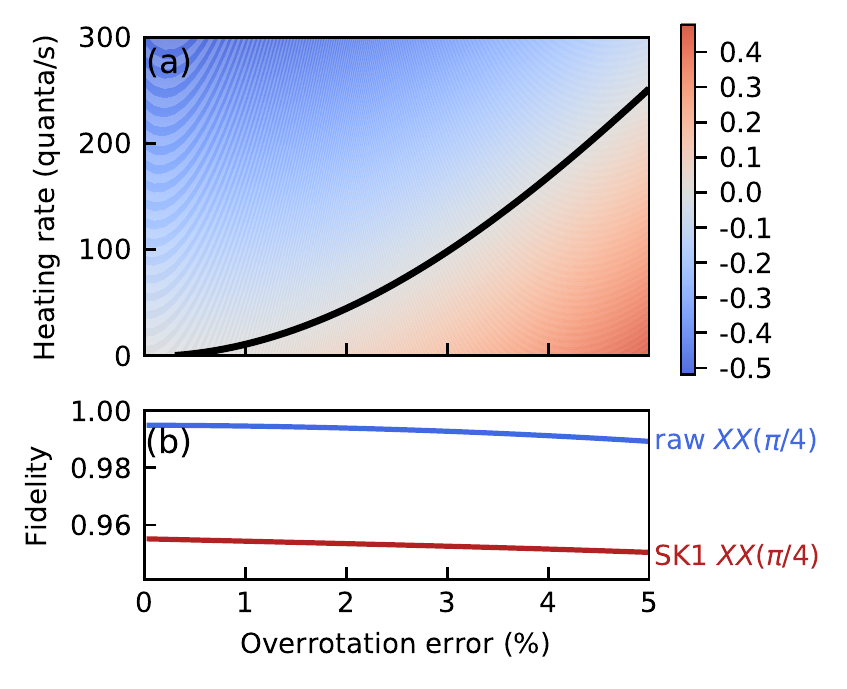}
		\caption[]{SK1 two-qubit gates' performance. (a) Simulation results of the average gate-fidelity improvement resulted from SK1 compensating pulses under different overrotation and stochastic error conditions. Warm (cool) color represents a fidelity improvement (degrading) comparing to the raw two-qubit gate. The diagonal curve in black is where the compensate pulses have a neutral impact. (b) Simulated fidelity of two-qubit SK1 compensating pulses and raw two-qubit gate. All dominant stochastic error sources in our experimental system are considered, including laser dephasing error, motional dephasing error, and motional heating.}
		\label{fig:sk1_viability}
	\end{center}
\end{figure}
Composite pulses developed for single-qubit gates to fix overrotations can be used to reduce overrotations in two-qubit gates using an isomorphism between one-qubit Pauli operators and a subgroup of two-qubit Pauli operators \cite{JonesPRA2003,TomitaNJP2010}. Previous calculations of hidden inverses built from composite two-qubit pulses were shown to greatly reduce circuit error in theory when the only error is gate overrotation \cite{murphy2019}. In practice, we have not seen an experimental advantage for these pulses. SK1 adds two additional $\pi$ MS gates resulting in a gate that is 3 times longer. 

We numerically consider the implementation of SK1 sequences for MS gate~\cite{murphy2019} using a simplified error model to understand why these methods do not provide an advantage. The average gate-fidelity difference is presented in Fig. \ref{fig:sk1_viability}(a). When we consider only overrotation error (coherent) and motional heating (stochastic), we find for an overrotation error that is around $1\%$ to $2\%$ that the motional heating rate would need to be as low as $20$ quanta per second for the SK1 sequence to improve gate performance. This is one order of magnitude lower than the heating rate in this system. When we consider all stochastic and coherent error sources in our system, SK1 sequences are predicted to severely limit the fidelity as shown in Fig. \ref{fig:sk1_viability}(b).

\subsection{Randomized compiling}
\begin{figure}[tp]
    \centering
    \includegraphics[width=\columnwidth]{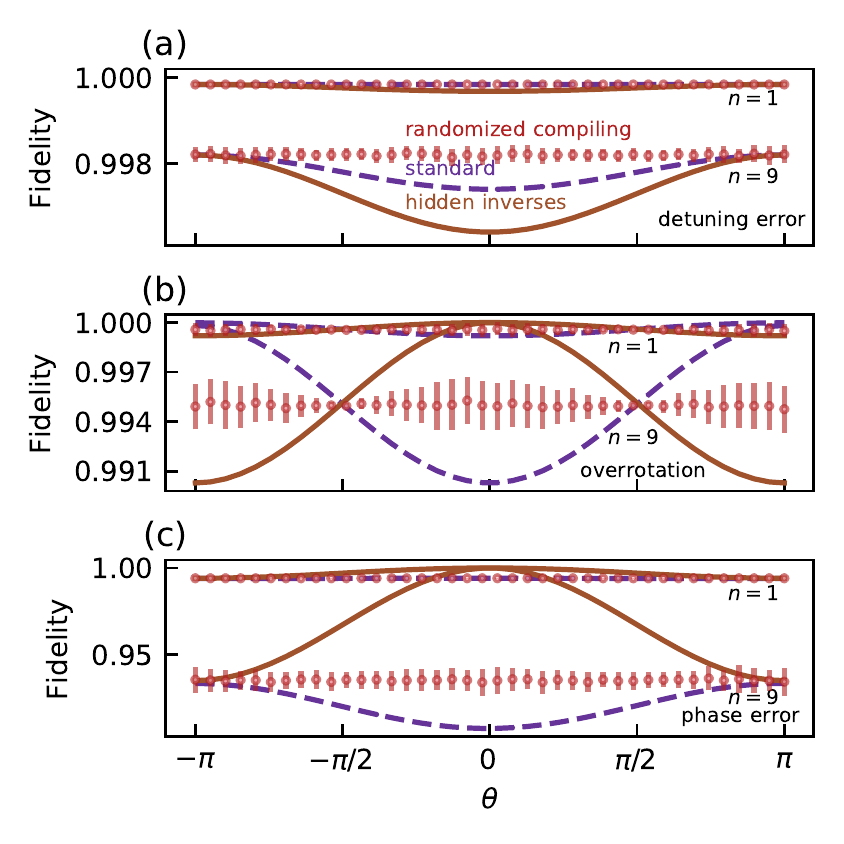}
    \caption{Randomized compiling performance. Final state fidelity comparison among hidden inverse configuration (brown), standard configuration (purple), and randomized compiling (red) executing circuits of Eq. \ref{eq:multi_qubit_cz} with different noise channels. (a) Detuning error ${\delta} = 1\%$ carrier Rabi frequency. (b) Two-qubit overrotation error $\epsilon_{2Q}=2\%$ and single-qubit overrotation $\epsilon_{1Q}=0.2\%$. (c) Phase misalignment $\phi_{\rm{diff}} = 3.5^\circ$ .}
    \label{fig:rc_compare}
\end{figure}
Randomized compiling (RC) \cite{wallman2016} is a protocol for converting coherent errors into stochastic errors. RC introduces independent random single-qubit gates into a circuit such that in the absence of noise, the overall ideal unitary remains the same. In the presence of noise, RC twirls the error channel into a stochastic Pauli channel. RC improves circuit results by preventing the worst-case cumulative errors and simplifies the prediction of algorithmic performance by reducing the complexity of the error model. 

In order to compare the performance of our hidden inverse protocol with randomized compiling, we numerically simulate each protocol on unitaries from Eq. \ref{eq:multi_qubit_cz} under three different noise models:  detuning error, overrotation, and phase misalignment. For the randomized compiling part, we sample 100 equivalent circuits for each value of $\theta$ (in $Z(\theta)$ from Eq. \ref{eq:multi_qubit_cz}) and take the average gate-fidelity of this ensemble. The average gate fidelity comparison is presented in Fig \ref{fig:rc_compare}. We find that hidden inverse configurations provide benefit over randomized compiling when the noise orientation of the error model is inverted with the inverse gate (Fig \ref{fig:rc_compare}(b), (c)). For errors that do not invert with the inverted gate controls, such as a detuning error, randomized compiling limits the coherent error accumulation providing a clear benefit over hidden inverses (Fig \ref{fig:rc_compare} (a)).

\subsection{Hardware-specific compilation}

Hidden inverses are developed in the context of a gate model of quantum computation. These gates need to be mapped onto a physical system and there are multiple software, and hardware layers between the user and the device. As a result, the operator of the quantum computer often prefers to compile any algorithm to the most hardware-efficient form to yield the highest overall fidelity.

Our running example circuit in this paper is the multiqubit-parity controlled Z rotation described in Eq. \ref{eq:multi_qubit_cz}. We show one way to map it to ion trap hardware but there are many hardware-specific ways to generate the same functionality. A clear example is that the base $n=2$ circuit of a $Z$ rotation by angle $\theta$ surrounded two $\CNOT$s which are composed of two $XX(\pi/4)$. This can be replaced by a single $XX(\theta/2)$ gate surrounded by single-qubit gates. Given that two-qubit gates are typically noisier than single-qubit gates,  this transformation is experimentally useful for quantum systems with Ising-type two-qubit couplings from nuclear magnetic resonance \cite{BrownPRL2006} to trapped ions\cite{NamNQuantumInf2019groundstate}. The cost here is that one needs to calibrate the two-qubit gate for multiple angles, which is inherently more error prone than the control of the $Z$ rotation, which in the experiment is only advancing a digital phase. We recognize that calibration may be less of a concern for near-term variational algorithms given the mismatch between algorithm performance at the ideal angle versus the programmed angle \cite{OMalleyPRX2016}.

For $n>2$, the additional $\CNOT$s could still benefit from hidden inverses, even if we change the internal primitive.  In some ion trap systems \cite{martinez2016compiling}, the natural multiqubit interaction is a global M\o lmer-S\o rensen. In this case, there could be a further reduction of the time complexity of the overall procedure. We have not considered this case in detail since our micro-mirror system is not compatible with a global M\o lmer-S\o rensen gate.

\subsection{Total optimization}

Various noise-adaptive compilers have been proposed recently in the literature. They include aggregation of multiple logical operations into larger units \cite{ShiASPLOS2019}, mapping and optimization of high-level quantum programs based on hardware specifications \cite{MuraliASPLOS2019}, and using machine learning and variational algorithms to develop noise-resilient circuits \cite{CincioPRXQuantum2021}, \cite{SharmaNJP2020}. While these methods outperform standard compilers for near-term devices with a few qubits and short depths, they are not expected to scale efficiently to be useful in large-scale fault-tolerant machines without truncation. Hidden inverses, on the other hand, take advantage of local optimization and can be efficiently included in compilers for larger quantum systems.


\section{Conclusions and outlook}
\label{sec:conclusions}
Slowly varying experimental noise sources can either be corrected by frequent calibrations or by introducing circuit-level protections such as composite pulses and hidden inverses.  By recognizing sets of gates that are self-adjoint, we can compile a circuit to cancel out coherent errors as long as the drift occurs at a timescale slower than the time between the two gates. We demonstrate a reduction of overrotations and phase misalignment for $\CNOT$ gates in an ion-trap system without changing the circuit length. Overall, these low-cost circuit compilation schemes provide a robust platform for reducing systematic error and have already been shown theoretically to provide an advantage for quantum chemistry circuits \cite{yeter2021benchmarking}.

Hidden inverses can be applied to any system where the gates are derived from flexible pulse control.  Hidden inverses can be further expanded to include gates that are only inverses on subspaces. From this viewpoint, we can reconsider the cancellation of coherent errors in stabilizer measurements by stabilizer slicing \cite{debroy2018} as a hidden inverse on the logical subspace. Hidden inverses also show the utility of having multiple versions of the same basic gate for improving circuit performance in the presence of systematic errors and suggest alternative user interfaces for quantum computers between a static set of gates and full pulse control.

\begin{acknowledgements}
The authors thank Erik Nielsen for helping with all pyGSTi-related queries. This work is supported by the Office of the Director of National Intelligence - Intelligence Advanced Research Projects Activity through ARO Contract No. W911NF-16-1-0082 (experimental implementation), National Science Foundation Expeditions in Computing Award 1730104 ($n$-qubit simulation), National Science Foundation STAQ Project  No. Phy-181891 (trapped-ion control sequences), the U.S. Department of Energy (DOE), Office of Advanced Scientific Computing Research award DE-SC0019294 (hidden inverse protocol), and DOE Basic Energy Sciences Award DE-0019449 (experimental analysis). S.M. is funded in part by a NSF QISE-NET fellowship (DMR-1747426).
\end{acknowledgements}

\appendix

\section{UNITARY $\CNOT$ ERROR MODEL}
\label{app:toymodel}
We consider the direct implementation of $\CNOT$ by a Hamiltonian \cite{BarnesPRA2017} in the context of hidden inverses.
\begin{eqnarray}
\CNOT_\epsilon&=&\exp\left(-i\frac{\epsilon}{2} \CNOT\right)\CNOT \\
\CNOT^\dagger_\epsilon&=&\CNOT\exp\left(i\frac{\epsilon}{2} \CNOT\right)
\end{eqnarray}
Here we consider the parity-controlled $Z$ rotation and calculate the average gate fidelity by calculating the entanglement fidelity.

The average gate fidelity between two unitary operations $U$ and $V$ on $n$ qubits is
\begin{eqnarray}
F(U,V)&=&\frac{2^{n}F_e(U,V)+1}{2^{n}+1}\\
&=&\frac{2^{n}\left(\frac{\left|\mathrm{Tr}\left[U^\dagger V\right]\right|^2}{4^{n}}\right)+1}{2^{n}+1}
\end{eqnarray} 
where $F_e(U,V)$ is the entanglement fidelity. 
In this  case, the ideal unitary for an $n-1$-qubit parity-controlled rotation of the target qubit $n$ is 
\begin{equation}
U=\exp\left(-i\frac{\theta}{2}\otimes_{j=1}^{n} Z_j\right)
\end{equation}
the actual unitary applied due to the systematic error is $V_\pm$
\begin{equation}
V_\pm=\left(\prod_{j=1}^{n-1} e^{\left(\pm i\frac{\epsilon}{2} \CNOT(j,n)\right)} \right)U\left(\prod_{k=1}^{n-1} e^{\left(- i\frac{\epsilon}{2} \CNOT(k,n)\right)} \right)
\end{equation}
where $V_+$ corresponds to hidden inverses and $V_-$ corresponds to the standard configuration.To calculate the entanglement fidelity we need to calculate $\mathrm{Tr}[U^\dagger V_\pm]$. First, we represent
\begin{eqnarray}
V_\pm&=&V_\pm U^\dagger U\\
&=&\left(\prod_{j=1}^{n-1} e^{\left(\pm i\frac{\epsilon}{2} \CNOT(j,n)\right)} \right)\left(\prod_{k=1}^{n-1} e^{\left(- i\frac{\epsilon}{2} \widetilde{\CNOT}(k,n)\right)}\right) \nonumber
\end{eqnarray}
where $\widetilde{\CNOT}=U {\CNOT}  U^\dagger$.
Next we rewrite $\CNOT(c,t)$ in terms of Pauli matrices and single qubit projectors
\begin{eqnarray}
\CNOT(c,t)=\frac{1}{2}(I+Z_c)+\frac{1}{2}(I-Z_c)X_t  \\
=\Pi_{0,c}+\Pi_{1,c}X_t 
\end{eqnarray}
where $\Pi_{\psi,a}$ projects qubit $a$ to the state $\psi$. 
This allows us to write 
\begin{eqnarray}
e^{-i\frac{\epsilon}{2}\CNOT(c,t)}=\left(\Pi_{0,c}e^{-i\frac{\epsilon}{2}}+\Pi_{1,c}e^{-i\frac{\epsilon}{2}X_t}\right)
\end{eqnarray}

We can write
\begin{equation}
e^{-i\frac{\epsilon}{2}\widetilde{\CNOT}(c,n)}=\left(\Pi_{0,c}e^{-i\frac{\epsilon}{2}}+\Pi_{1,c}e^{-i\frac{\epsilon}{2}\widetilde{X_{n}}}\right)
\end{equation}
where $\widetilde{X_{n}}=\cos(\theta)X_{n}+\sin(\theta)\left(\otimes_{j=1}^{n} Z_j\right)Y_{n}$ and we define $\widetilde{X}(\theta)=\cos(\theta)X+\sin(\theta)Y$

Due to the projectors, each string of bits on the first $n$ qubits will generate a residual unitary operation on the target qubit that depends only on the Haming weight $w$, the number of $1$'s in the bit string. The parity of the bit string determines the sign of $\theta$ in $\widetilde{X}(\theta)$ 

Putting it all together we have
\begin{eqnarray}
\nonumber
    \mathrm{Tr}[U^\dagger V_{\pm}]&=&\mathrm{Tr}\left[\prod_{j=1}^{n-1} e^{\pm i\frac{\epsilon}{2}\CNOT(j,n)} \prod_{j=1}^{n-1} e^{- i\frac{\epsilon}{2}\widetilde{\CNOT}(j,n)}\right]\\ \nonumber
    &=&\sum_{w=0}^{n-1} {n-1\choose w} e^{-i\frac{(n-1-w)\epsilon}{2}(-1\pm 1)}\mathrm{Tr}_2[e^{\pm i \frac{w\epsilon}{2} X}e^{-i\frac{w\epsilon}{2}\widetilde{X}(-1^w\theta)}]\\
    &=&\sum_{w=0}^{n-1} {n-1\choose w} e^{-i\frac{(n-1-w)\epsilon}{2}(-1\pm 1)}B_\pm(w,\theta) 
\end{eqnarray}
where $\mathrm{Tr}_2$ is the trace over a two-dimensional space and
\begin{equation}
    B_\pm(w,\theta)=2\left(\cos\left(w\frac{\epsilon}{2}\right)^2\pm \cos(\theta)\sin\left(w\frac{\epsilon}{2}\right)^2\right)
\end{equation}
We note that that $B_+(w,\theta)=B_-(w,\theta+\pi)$ this leads to the maximum fidelities happening at different $\theta$ and the fidelities oscillating $\pi$ out of phase.
We find for hidden inverses $F_e(U,V_+)|_{\theta=0}$=1 and $F_e(U,V_+)|_{\theta=\pi}=\cos((n-1)\epsilon/2)\cos(\epsilon/2)^{2(n-1)}$ and for the standard configuration $F_e(U,V_-)|_{\theta=0}=\frac{1}{4}\left(1+2\cos((n-1)\epsilon)\cos(\epsilon)^{n-1}+\cos(\epsilon)^{2(n-1)}\right)$ and  $F_e(U,V_-)|_{\theta=\pi}=\cos(\epsilon/2)^{2(n-1)}$ using the mathematical identity
\begin{equation}
  \sum_{w=0}^{n-1} {n-1\choose w} e^{-iw\epsilon}= e^{-i(n-1)\epsilon/2}\left[2\cos(\epsilon/2)\right]^{n-1}  
\end{equation}.

\section{MS GATE ERROR MODEL}
\label{sec:errormodel}

The MS gate error model can be found in the Supplementary Material of Ref \cite{wang2020}. We present it here for convenience. We make some updates for the error model to simulate quantum circuits efficiently.

The Hamiltonian of the MS evolution of the $j$th motional mode with no modulation is written as \cite{sorensen1999,Sorensen2000,hayes2012remote}
\begin{widetext}
\begin{eqnarray}
\label{eq1}\hat{H}(t)_{j,\rm{MS}} = \frac{i}{2} \sum_{n=1,2} \eta_{j}^{(n)}\hat{\sigma}_{+}^{(n)}\left(\Omega_{r}^{(n)} \hat{a}_j e^{i \phi_{r}-i \delta_{j,r}^{(n)}t}+\Omega_{b}^{(n)}\hat{a}_j^{\dag}e^{i \phi_{b}-i \delta_{j,b}^{(n)}t} \right) + h.c.
\end{eqnarray}
\end{widetext}
where $\Omega_{r}^{(1)}$, $\Omega_{b}^{(1)}$, $\Omega_{r}^{(2)}$, and $\Omega_{b}^{(2)}$ are the Rabi frequencies of red and blue sideband transitions for the two target ions, $\delta_{j,r}^{(1)}$, $\delta_{j,b}^{(1)}$, $\delta_{j,r}^{(2)}$, and $\delta_{j,b}^{(2)}$ are the detunings for the $j$th motional mode, $\phi_r$ and $\phi_b$ are the laser phases of the red and blue tone, respectively. With the expansion in Eq. (\ref{eq1}), we can simulate the number of error mechanisms: power imbalance on two target ions, power imbalance on red and blue tones, and detuning imbalance due to Stark shift. For the full MS evolution, the modes are sequentially simulated to minimize the computing resource. We save only the spin-state result for the next round of simulation. The Hamiltonian of different modes commute when $\Omega_{r}^{(1)} = \Omega_{b}^{(1)}$ and $\Omega_{r}^{(2)} = \Omega_{b}^{(2)}$, which is a reasonable assumption in the MS gate. For the evolution of discrete segments in FM gates, we sequentially simulate every segment to obtain the final state.

We use a master equation \cite{gardiner2004quantum} to simulate an open-quantum system considering multiple dissipative error mechanisms: motional heating, motional dephasing, and laser dephasing. The master equation is written in Lindblad form~\cite{lindblad1976generators}
\begin{equation*}
    \frac{d\hat{\rho}}{d t} = \frac{1}{i \hbar}[\hat{H},\hat{\rho}]+\sum_j\left( \hat{L}_j\hat{\rho} \hat{L}_j^{\dag} -\frac{1}{2} \hat{L}_j^{\dag} \hat{L}_j \hat{\rho} -\frac{1}{2} \hat{\rho} \hat{L}_j^{\dag} \hat{L}_j \right),
\end{equation*}
where $\rho$ is the density matrix of the system, $H$ is the Hamiltonian of the MS gate, $\hat{L}_j$ is the Lindblad operator for the $j$th decoherence process. The motional dephasing can be described by the Lindblad operator of the form $\hat{L}_m = \sqrt{\frac{2}{\tau_m}}\hat{a}^{\dag}\hat{a}$, where $\tau_m$ is the motional coherence time. The anomalous heating can be described by $\hat{L}_{+} = \sqrt{\Gamma}\hat{a}^{\dag}$ and $\hat{L}_{-} = \sqrt{\Gamma}\hat{a}$, where $\Gamma$ is the heating rate. For these two operators, we sequentially simulate the evolution of each mode, then combine them to obtain the final state. The master-equation simulations represent the full density-matrix representation for a truncated state space of two qubits and one motional mode truncated to the first 13 Fock states ($n \leq 12$). The laser dephasing can be described by the Lindblad operator of $\hat{L}_l = \sqrt{1/\tau_l}(\hat{\sigma}_{z}^{(1)}+\hat{\sigma}_{z}^{(2)})$, where $\tau_l$ is the laser coherence time. For this Lindblad operator, we perform a full master-equation simulation with all motional modes and spin states included. We truncate the far off-resonance motional modes, which have a smaller motional excitation, to smaller Fock states to save on computational resources. For the stochastic noise, we also combine the simulation with the Monte Carlo method. The simulations are performed using QuTip \cite{johansson2012qutip}.

To avoid solving master equations whenever we encounter a MS gate in the quantum circuit, we calculate the Pauli transfer matrices (PTMs) before simulating the circuit. The PTM is represented as:
\begin{equation}
    (R_\Lambda)_{ij} = \frac{1}{d}\rm{Tr}\{P_i\Lambda(P_j)\},
\end{equation}
where $P_i$ is the Pauli basis, $d=2^n, n$ is the number of qubits, and $\Lambda$ is the linear map \cite{greenbaum2015introduction}. $\Lambda(P_j)$ is equivalent to applying the master-equation simulation on Pauli basis $P_j$. Single-qubit gates suffer from negligible stochastic noises. Therefore, we represent them with corresponding quantum operation matrices subject to minor coherent errors. In superoperator formalism, a quantum circuit comprised of quantum maps (the M\o lmer-S\o rensen gates and the single-qubit rotations) is equivalent to matrix multiplication of the corresponding PTMs and can be calculated efficiently.

\section{GATE SET TOMOGRAPHY FOR SINGLE-QUBIT GATES}
\label{sec:gst}

\begin{figure}[tp]
	\begin{center}
		\includegraphics[width=\columnwidth]{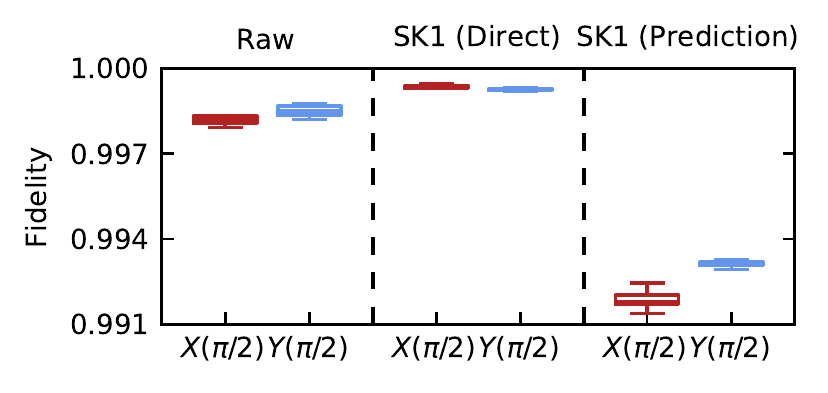}
		\caption[]{Average gate-fidelity comparison. We compare the average gate fidelity of three instances of $X(\pi/2)$ (red) and $Y(\pi/2)$ (blue) gates. The three cases considered are direct characterization of raw gates, direct characterization of SK1 gates, and predicted characterization of SK1 gates based on the results from GST on the raw gates. The boxplot displays the minimum, the maximum, the sample median, and the first and third quartiles of the dataset.}
		\label{fig:Fidelity GxGy}
	\end{center}
\end{figure}

We design an experiment to measure the performance of SK1 gates and test how well GST predicts their performance. The experiment serves as a preliminary systematic error characterization. First, we run GST on a gate set composed of the SK1 compiled gates $\{X_{\rm{SK1}}(\pi/2)$, $Y_{\rm{SK1}}(\pi/2)\}$, followed by an experiment where we run GST on a gate set comprised of the raw gates that generate SK1 sequences $\{X(\pi/2)$, $Y(\pi/2)$, ${\rm{SK1}}^+_{X}(2\pi)$, ${\rm{SK1}}^-_{X}(2\pi)$, ${\rm{SK1}}^+_{Y}(2\pi)$, ${\rm{SK1}}^-_{Y}(2\pi)\}$. GST produces a completely positive trace-preserving map for each gate, represented as a PTM.

We calculate the fidelity of the SK1 gates and the raw gates from these PTMs. The PTMs allow us to calculate any fidelity and we choose the average gate fidelity, $\mathcal{F}(U,\mathcal{E})=\int d\psi \bra{\psi}U^\dagger\mathcal{E}(\ket{\psi}\bra{\psi}) U \ket{\psi}$ where $U$ is the ideal gate and $\mathcal{E}$ is the actual gate \cite{Nielsen1996}.    From the GST PTMs, we calculate a fidelity for SK1 $X(\pi/2)$ and SK1 $Y(\pi/2)$ of $0.999~36 (5)$ and $0.999~27 (3)$, respectively, while the fidelity for raw-compiled $X(\pi/2)$ is $0.9982 (1)$ and for $Y(\pi/2)$ is $0.9985(2)$. We see a clear improvement in fidelity due to the SK1 composite pulses. Also, smaller error bars in the calculated fidelity of SK1 gates indicate that the gates are more uniform. The estimated error generator for each gate, which is a Lindbladian type operator that acts after the ideal gate ($G=e^{\mathbb{L}} G_{0}$), describes how the gate is failing to match the target. Specifically, the Hamiltonian projection of this error generator produces the coherent part of the error. We find that SK1 turns approximately $1\%$ overrotation into approximately $0.01\%$ overrotation as expected.

We then combine the PTMs obtained from the raw-pulse GST to construct SK1 $X(\pi/2)$ and $Y(\pi/2)$ gate PTMs. Notice that the constructed PTMs are significantly different from the direct SK1 gate PTMs obtained from composite pulse GST. The fidelity for the predicted SK1 $X(\pi/2)$ gate is $0.9917 (4)$, and for the $Y(\pi/2)$ gate it is $0.9931 (2)$. Fig. \ref{fig:Fidelity GxGy} contains box plots of the calculated fidelities. It indicates that raw-pulse GST predicts SK1 composite pulses degrade gate fidelities. This result contradicts the experiment, where SK1 does improve the gate performance. This discrepancy can be explained by an overrotation error that is slowly varying. The raw pulse GST averages over the time-varying overrotations, yielding a PTM that describes average raw pulses for which SK1 would not be useful. Simulations readily reproduce this behavior. 

We use pyGSTi (version 0.9.9.1) \cite{Nielsen2020} for all GST-related works. This section explains the experimental design and data analysis for characterizing SK1 gates and their elementary rotations. 


\textit{Experiment design}---The experimental circuits are generated by pyGSTi's fiducial and germ selection algorithms. Fiducial sequences are used to prepare and measure an informationally complete set of operations. Germs are designed to amplify all possible gate errors. Given a set of operations (also called the gate set), we use the algorithms to generate the appropriate fiducials and germs. Our gate sets are $\{X_{\rm{SK1}}(\pi/2),Y_{\rm{SK1}}(\pi/2)\}$ and $\{X(\pi/2)$, $Y(\pi/2)$, ${\rm{SK1}}^+_{X}(2\pi)$, ${\rm{SK1}}^-_{X}(2\pi)$, ${\rm{SK1}}^+_{Y}(2\pi)$, ${\rm{SK1}}^-_{Y}(2\pi)\}$. Fiducials and germs in hand, we choose the length of the experiments (number of times to repeat each germ between fiducial pairs) as $L=256$ and $L=32$, respectively. The experiment lengths are different for two gate sets because raw gates are noisier than composite pulse gates, and raw gates reach a similar noise level as composite pulse gates with less noise amplification.

\textit{Data analysis}.---We run standard GST as implemented in pyGSTi. Results of the gate set $\{X_{{\rm{SK1}}}(\pi/2), Y_{{\rm{SK1}}}(\pi/2)\}$ follow directly from the output provided by GST (other than error bars, which we discuss next). For the gate set $\{X(\pi/2)$, $Y(\pi/2)$, ${\rm{SK1}}^+_{X}(2\pi)$, ${\rm{SK1}}^-_{X}(2\pi)$, ${\rm{SK1}}^+_{Y}(2\pi)$, ${\rm{SK1}}^-_{Y}(2\pi)\}$, we get the PTMs for the elementary rotations directly from GST. We calculate the predicted SK1 gate PTMs through matrix multiplication of the elementary rotation PTMs,   
$$
R_{{\rm{SK1}}}(\theta, \phi)={\rm{SK1}}^-_{R}(2\pi) {\rm{SK1}}^+_{R}(2\pi) R(\pi/2)  ,
$$
where $R \in  \{X,Y\}$. 
To generate the error bars on the calculated fidelity metrics, we use a nonparametric bootstrapping technique from pyGSTi. We take the final estimate from running standard GST as the target model for generating nonparametric bootstrapping samples and then run gauge optimization on these raw bootstrapped models to generate our final set of models. Error bars are calculated from the standard deviation of average gate-fidelity metrics on the set. GST provides information on ``Goodness of fit,'' i.e., how well GST estimates the fit to characterize the data, to provide confidence in the data analysis. A rating scale from 1 to 5 summarizes various statistical measures. For both gate sets, the experiments receive a score higher than 4 indicating a good fit. 

The noise in the PTMs can be better understood using projections of the gate error generators. These are Linbladian-like operators generated by projecting the error generator into some subspace. We are primarily concerned with the Hamiltonian projection, which produces the coherent error. We use built-in pyGSTi functions to calculate these projections and deduce the amount of overrotation error.


\bibliography{References}

\begin{thebibliography}{68}%
\makeatletter
\providecommand \@ifxundefined [1]{%
 \@ifx{#1\undefined}
}%
\providecommand \@ifnum [1]{%
 \ifnum #1\expandafter \@firstoftwo
 \else \expandafter \@secondoftwo
 \fi
}%
\providecommand \@ifx [1]{%
 \ifx #1\expandafter \@firstoftwo
 \else \expandafter \@secondoftwo
 \fi
}%
\providecommand \natexlab [1]{#1}%
\providecommand \enquote  [1]{``#1''}%
\providecommand \bibnamefont  [1]{#1}%
\providecommand \bibfnamefont [1]{#1}%
\providecommand \citenamefont [1]{#1}%
\providecommand \href@noop [0]{\@secondoftwo}%
\providecommand \href [0]{\begingroup \@sanitize@url \@href}%
\providecommand \@href[1]{\@@startlink{#1}\@@href}%
\providecommand \@@href[1]{\endgroup#1\@@endlink}%
\providecommand \@sanitize@url [0]{\catcode `\\12\catcode `\$12\catcode
  `\&12\catcode `\#12\catcode `\^12\catcode `\_12\catcode `\%12\relax}%
\providecommand \@@startlink[1]{}%
\providecommand \@@endlink[0]{}%
\providecommand \url  [0]{\begingroup\@sanitize@url \@url }%
\providecommand \@url [1]{\endgroup\@href {#1}{\urlprefix }}%
\providecommand \urlprefix  [0]{URL }%
\providecommand \Eprint [0]{\href }%
\providecommand \doibase [0]{https://doi.org/}%
\providecommand \selectlanguage [0]{\@gobble}%
\providecommand \bibinfo  [0]{\@secondoftwo}%
\providecommand \bibfield  [0]{\@secondoftwo}%
\providecommand \translation [1]{[#1]}%
\providecommand \BibitemOpen [0]{}%
\providecommand \bibitemStop [0]{}%
\providecommand \bibitemNoStop [0]{.\EOS\space}%
\providecommand \EOS [0]{\spacefactor3000\relax}%
\providecommand \BibitemShut  [1]{\csname bibitem#1\endcsname}%
\let\auto@bib@innerbib\@empty
\bibitem [{\citenamefont {Hahn}(1950)}]{hahn1950}%
  \BibitemOpen
  \bibfield  {author} {\bibinfo {author} {\bibfnamefont {E.~L.}\ \bibnamefont
  {Hahn}},\ }\bibfield  {title} {\bibinfo {title} {Spin echoes},\ }\href
  {https://doi.org/10.1103/PhysRev.80.580} {\bibfield  {journal} {\bibinfo
  {journal} {Phys. Rev.}\ }\textbf {\bibinfo {volume} {80}},\ \bibinfo {pages}
  {580} (\bibinfo {year} {1950})}\BibitemShut {NoStop}%
\bibitem [{\citenamefont {Viola}\ \emph {et~al.}(1999)\citenamefont {Viola},
  \citenamefont {Knill},\ and\ \citenamefont {Lloyd}}]{viola1999}%
  \BibitemOpen
  \bibfield  {author} {\bibinfo {author} {\bibfnamefont {L.}~\bibnamefont
  {Viola}}, \bibinfo {author} {\bibfnamefont {E.}~\bibnamefont {Knill}},\ and\
  \bibinfo {author} {\bibfnamefont {S.}~\bibnamefont {Lloyd}},\ }\bibfield
  {title} {\bibinfo {title} {Dynamical decoupling of open quantum systems},\
  }\href {https://doi.org/10.1103/PhysRevLett.82.2417} {\bibfield  {journal}
  {\bibinfo  {journal} {Phys. Rev. Lett.}\ }\textbf {\bibinfo {volume} {82}},\
  \bibinfo {pages} {2417} (\bibinfo {year} {1999})}\BibitemShut {NoStop}%
\bibitem [{\citenamefont {Biercuk}\ \emph {et~al.}(2009)\citenamefont
  {Biercuk}, \citenamefont {Uys}, \citenamefont {VanDevender}, \citenamefont
  {Shiga}, \citenamefont {Itano},\ and\ \citenamefont
  {Bollinger}}]{biercuk2009optimized}%
  \BibitemOpen
  \bibfield  {author} {\bibinfo {author} {\bibfnamefont {M.~J.}\ \bibnamefont
  {Biercuk}}, \bibinfo {author} {\bibfnamefont {H.}~\bibnamefont {Uys}},
  \bibinfo {author} {\bibfnamefont {A.~P.}\ \bibnamefont {VanDevender}},
  \bibinfo {author} {\bibfnamefont {N.}~\bibnamefont {Shiga}}, \bibinfo
  {author} {\bibfnamefont {W.~M.}\ \bibnamefont {Itano}},\ and\ \bibinfo
  {author} {\bibfnamefont {J.~J.}\ \bibnamefont {Bollinger}},\ }\bibfield
  {title} {\bibinfo {title} {Optimized dynamical decoupling in a model quantum
  memory},\ }\href@noop {} {\bibfield  {journal} {\bibinfo  {journal} {Nature}\
  }\textbf {\bibinfo {volume} {458}},\ \bibinfo {pages} {996} (\bibinfo {year}
  {2009})}\BibitemShut {NoStop}%
\bibitem [{\citenamefont {Qi}\ \emph {et~al.}(2017)\citenamefont {Qi},
  \citenamefont {Dowling},\ and\ \citenamefont {Viola}}]{qi2017optimal}%
  \BibitemOpen
  \bibfield  {author} {\bibinfo {author} {\bibfnamefont {H.}~\bibnamefont
  {Qi}}, \bibinfo {author} {\bibfnamefont {J.~P.}\ \bibnamefont {Dowling}},\
  and\ \bibinfo {author} {\bibfnamefont {L.}~\bibnamefont {Viola}},\ }\bibfield
   {title} {\bibinfo {title} {Optimal digital dynamical decoupling for general
  decoherence via walsh modulation},\ }\href@noop {} {\bibfield  {journal}
  {\bibinfo  {journal} {Quantum Information Processing}\ }\textbf {\bibinfo
  {volume} {16}},\ \bibinfo {pages} {1} (\bibinfo {year} {2017})}\BibitemShut
  {NoStop}%
\bibitem [{\citenamefont {Quiroz}\ and\ \citenamefont
  {Lidar}(2013)}]{quiroz2013}%
  \BibitemOpen
  \bibfield  {author} {\bibinfo {author} {\bibfnamefont {G.}~\bibnamefont
  {Quiroz}}\ and\ \bibinfo {author} {\bibfnamefont {D.~A.}\ \bibnamefont
  {Lidar}},\ }\bibfield  {title} {\bibinfo {title} {Optimized dynamical
  decoupling via genetic algorithms},\ }\href
  {https://doi.org/10.1103/PhysRevA.88.052306} {\bibfield  {journal} {\bibinfo
  {journal} {Phys. Rev. A}\ }\textbf {\bibinfo {volume} {88}},\ \bibinfo
  {pages} {052306} (\bibinfo {year} {2013})}\BibitemShut {NoStop}%
\bibitem [{\citenamefont {Lidar}(2014)}]{lidar2014review}%
  \BibitemOpen
  \bibfield  {author} {\bibinfo {author} {\bibfnamefont {D.~A.}\ \bibnamefont
  {Lidar}},\ }\bibfield  {title} {\bibinfo {title} {Review of decoherence free
  subspaces, noiseless subsystems, and dynamical decoupling},\ }\href@noop {}
  {\bibfield  {journal} {\bibinfo  {journal} {Adv. Chem. Phys}\ }\textbf
  {\bibinfo {volume} {154}},\ \bibinfo {pages} {295} (\bibinfo {year}
  {2014})}\BibitemShut {NoStop}%
\bibitem [{\citenamefont {Zeng}\ \emph {et~al.}(2018)\citenamefont {Zeng},
  \citenamefont {Deng}, \citenamefont {Russo},\ and\ \citenamefont
  {Barnes}}]{zeng2018general}%
  \BibitemOpen
  \bibfield  {author} {\bibinfo {author} {\bibfnamefont {J.}~\bibnamefont
  {Zeng}}, \bibinfo {author} {\bibfnamefont {X.-H.}\ \bibnamefont {Deng}},
  \bibinfo {author} {\bibfnamefont {A.}~\bibnamefont {Russo}},\ and\ \bibinfo
  {author} {\bibfnamefont {E.}~\bibnamefont {Barnes}},\ }\bibfield  {title}
  {\bibinfo {title} {General solution to inhomogeneous dephasing and smooth
  pulse dynamical decoupling},\ }\href@noop {} {\bibfield  {journal} {\bibinfo
  {journal} {New Journal of Physics}\ }\textbf {\bibinfo {volume} {20}},\
  \bibinfo {pages} {033011} (\bibinfo {year} {2018})}\BibitemShut {NoStop}%
\bibitem [{\citenamefont {Merrill}\ and\ \citenamefont
  {Brown}(2014)}]{merrill2014progress}%
  \BibitemOpen
  \bibfield  {author} {\bibinfo {author} {\bibfnamefont {J.~T.}\ \bibnamefont
  {Merrill}}\ and\ \bibinfo {author} {\bibfnamefont {K.~R.}\ \bibnamefont
  {Brown}},\ }\bibfield  {title} {\bibinfo {title} {Progress in compensating
  pulse sequences for quantum computation},\ }\href@noop {} {\bibfield
  {journal} {\bibinfo  {journal} {Adv. Chem. Phys.}\ }\textbf {\bibinfo
  {volume} {154}},\ \bibinfo {pages} {241} (\bibinfo {year}
  {2014})}\BibitemShut {NoStop}%
\bibitem [{\citenamefont {Low}\ \emph {et~al.}(2014)\citenamefont {Low},
  \citenamefont {Yoder},\ and\ \citenamefont {Chuang}}]{low2014}%
  \BibitemOpen
  \bibfield  {author} {\bibinfo {author} {\bibfnamefont {G.~H.}\ \bibnamefont
  {Low}}, \bibinfo {author} {\bibfnamefont {T.~J.}\ \bibnamefont {Yoder}},\
  and\ \bibinfo {author} {\bibfnamefont {I.~L.}\ \bibnamefont {Chuang}},\
  }\bibfield  {title} {\bibinfo {title} {Optimal arbitrarily accurate composite
  pulse sequences},\ }\href {https://doi.org/10.1103/PhysRevA.89.022341}
  {\bibfield  {journal} {\bibinfo  {journal} {Phys. Rev. A}\ }\textbf {\bibinfo
  {volume} {89}},\ \bibinfo {pages} {022341} (\bibinfo {year}
  {2014})}\BibitemShut {NoStop}%
\bibitem [{\citenamefont {Low}\ \emph {et~al.}(2016)\citenamefont {Low},
  \citenamefont {Yoder},\ and\ \citenamefont {Chuang}}]{low2016}%
  \BibitemOpen
  \bibfield  {author} {\bibinfo {author} {\bibfnamefont {G.~H.}\ \bibnamefont
  {Low}}, \bibinfo {author} {\bibfnamefont {T.~J.}\ \bibnamefont {Yoder}},\
  and\ \bibinfo {author} {\bibfnamefont {I.~L.}\ \bibnamefont {Chuang}},\
  }\bibfield  {title} {\bibinfo {title} {Methodology of resonant equiangular
  composite quantum gates},\ }\href {https://doi.org/10.1103/PhysRevX.6.041067}
  {\bibfield  {journal} {\bibinfo  {journal} {Phys. Rev. X}\ }\textbf {\bibinfo
  {volume} {6}},\ \bibinfo {pages} {041067} (\bibinfo {year}
  {2016})}\BibitemShut {NoStop}%
\bibitem [{\citenamefont {Mount}\ \emph {et~al.}(2015)\citenamefont {Mount},
  \citenamefont {Kabytayev}, \citenamefont {Crain}, \citenamefont {Harper},
  \citenamefont {Baek}, \citenamefont {Vrijsen}, \citenamefont {Flammia},
  \citenamefont {Brown}, \citenamefont {Maunz},\ and\ \citenamefont
  {Kim}}]{MountPRA2015}%
  \BibitemOpen
  \bibfield  {author} {\bibinfo {author} {\bibfnamefont {E.}~\bibnamefont
  {Mount}}, \bibinfo {author} {\bibfnamefont {C.}~\bibnamefont {Kabytayev}},
  \bibinfo {author} {\bibfnamefont {S.}~\bibnamefont {Crain}}, \bibinfo
  {author} {\bibfnamefont {R.}~\bibnamefont {Harper}}, \bibinfo {author}
  {\bibfnamefont {S.-Y.}\ \bibnamefont {Baek}}, \bibinfo {author}
  {\bibfnamefont {G.}~\bibnamefont {Vrijsen}}, \bibinfo {author} {\bibfnamefont
  {S.~T.}\ \bibnamefont {Flammia}}, \bibinfo {author} {\bibfnamefont {K.~R.}\
  \bibnamefont {Brown}}, \bibinfo {author} {\bibfnamefont {P.}~\bibnamefont
  {Maunz}},\ and\ \bibinfo {author} {\bibfnamefont {J.}~\bibnamefont {Kim}},\
  }\bibfield  {title} {\bibinfo {title} {Error compensation of single-qubit
  gates in a surface-electrode ion trap using composite pulses},\ }\href
  {https://doi.org/10.1103/PhysRevA.92.060301} {\bibfield  {journal} {\bibinfo
  {journal} {Phys. Rev. A}\ }\textbf {\bibinfo {volume} {92}},\ \bibinfo
  {pages} {060301(R)} (\bibinfo {year} {2015})}\BibitemShut {NoStop}%
\bibitem [{\citenamefont {Khodjasteh}\ and\ \citenamefont
  {Viola}(2009)}]{KhodjastehPRL2009}%
  \BibitemOpen
  \bibfield  {author} {\bibinfo {author} {\bibfnamefont {K.}~\bibnamefont
  {Khodjasteh}}\ and\ \bibinfo {author} {\bibfnamefont {L.}~\bibnamefont
  {Viola}},\ }\bibfield  {title} {\bibinfo {title} {Dynamically error-corrected
  gates for universal quantum computation},\ }\href
  {https://doi.org/10.1103/PhysRevLett.102.080501} {\bibfield  {journal}
  {\bibinfo  {journal} {Phys. Rev. Lett.}\ }\textbf {\bibinfo {volume} {102}},\
  \bibinfo {pages} {080501} (\bibinfo {year} {2009})}\BibitemShut {NoStop}%
\bibitem [{\citenamefont {Buterakos}\ \emph {et~al.}(2021)\citenamefont
  {Buterakos}, \citenamefont {Das~Sarma},\ and\ \citenamefont
  {Barnes}}]{ButerakosPRXQ2021}%
  \BibitemOpen
  \bibfield  {author} {\bibinfo {author} {\bibfnamefont {D.}~\bibnamefont
  {Buterakos}}, \bibinfo {author} {\bibfnamefont {S.}~\bibnamefont
  {Das~Sarma}},\ and\ \bibinfo {author} {\bibfnamefont {E.}~\bibnamefont
  {Barnes}},\ }\bibfield  {title} {\bibinfo {title} {Geometrical formalism for
  dynamically corrected gates in multiqubit systems},\ }\href
  {https://doi.org/10.1103/PRXQuantum.2.010341} {\bibfield  {journal} {\bibinfo
   {journal} {PRX Quantum}\ }\textbf {\bibinfo {volume} {2}},\ \bibinfo {pages}
  {010341} (\bibinfo {year} {2021})}\BibitemShut {NoStop}%
\bibitem [{\citenamefont {Wallman}\ and\ \citenamefont
  {Emerson}(2016)}]{wallman2016}%
  \BibitemOpen
  \bibfield  {author} {\bibinfo {author} {\bibfnamefont {J.~J.}\ \bibnamefont
  {Wallman}}\ and\ \bibinfo {author} {\bibfnamefont {J.}~\bibnamefont
  {Emerson}},\ }\bibfield  {title} {\bibinfo {title} {Noise tailoring for
  scalable quantum computation via randomized compiling},\ }\href
  {https://doi.org/10.1103/PhysRevA.94.052325} {\bibfield  {journal} {\bibinfo
  {journal} {Phys. Rev. A}\ }\textbf {\bibinfo {volume} {94}},\ \bibinfo
  {pages} {052325} (\bibinfo {year} {2016})}\BibitemShut {NoStop}%
\bibitem [{\citenamefont {Lidar}\ \emph {et~al.}(1998)\citenamefont {Lidar},
  \citenamefont {Chuang},\ and\ \citenamefont {Whaley}}]{lidar1998}%
  \BibitemOpen
  \bibfield  {author} {\bibinfo {author} {\bibfnamefont {D.~A.}\ \bibnamefont
  {Lidar}}, \bibinfo {author} {\bibfnamefont {I.~L.}\ \bibnamefont {Chuang}},\
  and\ \bibinfo {author} {\bibfnamefont {K.~B.}\ \bibnamefont {Whaley}},\
  }\bibfield  {title} {\bibinfo {title} {Decoherence-free subspaces for quantum
  computation},\ }\href {https://doi.org/10.1103/PhysRevLett.81.2594}
  {\bibfield  {journal} {\bibinfo  {journal} {Phys. Rev. Lett.}\ }\textbf
  {\bibinfo {volume} {81}},\ \bibinfo {pages} {2594} (\bibinfo {year}
  {1998})}\BibitemShut {NoStop}%
\bibitem [{\citenamefont {Kwiat}\ \emph {et~al.}(2000)\citenamefont {Kwiat},
  \citenamefont {Berglund}, \citenamefont {Altepeter},\ and\ \citenamefont
  {White}}]{kwiat2000}%
  \BibitemOpen
  \bibfield  {author} {\bibinfo {author} {\bibfnamefont {P.~G.}\ \bibnamefont
  {Kwiat}}, \bibinfo {author} {\bibfnamefont {A.~J.}\ \bibnamefont {Berglund}},
  \bibinfo {author} {\bibfnamefont {J.~B.}\ \bibnamefont {Altepeter}},\ and\
  \bibinfo {author} {\bibfnamefont {A.~G.}\ \bibnamefont {White}},\ }\bibfield
  {title} {\bibinfo {title} {Experimental verification of decoherence-free
  subspaces},\ }\href {https://doi.org/10.1126/science.290.5491.498} {\bibfield
   {journal} {\bibinfo  {journal} {Science}\ }\textbf {\bibinfo {volume}
  {290}},\ \bibinfo {pages} {498} (\bibinfo {year} {2000})}\BibitemShut
  {NoStop}%
\bibitem [{\citenamefont {Debroy}\ \emph {et~al.}(2021)\citenamefont {Debroy},
  \citenamefont {Egan}, \citenamefont {Noel}, \citenamefont {Risinger},
  \citenamefont {Zhu}, \citenamefont {Biswas}, \citenamefont {Cetina},
  \citenamefont {Monroe},\ and\ \citenamefont {Brown}}]{debroy2021optimizing}%
  \BibitemOpen
  \bibfield  {author} {\bibinfo {author} {\bibfnamefont {D.~M.}\ \bibnamefont
  {Debroy}}, \bibinfo {author} {\bibfnamefont {L.}~\bibnamefont {Egan}},
  \bibinfo {author} {\bibfnamefont {C.}~\bibnamefont {Noel}}, \bibinfo {author}
  {\bibfnamefont {A.}~\bibnamefont {Risinger}}, \bibinfo {author}
  {\bibfnamefont {D.}~\bibnamefont {Zhu}}, \bibinfo {author} {\bibfnamefont
  {D.}~\bibnamefont {Biswas}}, \bibinfo {author} {\bibfnamefont
  {M.}~\bibnamefont {Cetina}}, \bibinfo {author} {\bibfnamefont
  {C.}~\bibnamefont {Monroe}},\ and\ \bibinfo {author} {\bibfnamefont {K.~R.}\
  \bibnamefont {Brown}},\ }\bibfield  {title} {\bibinfo {title} {Optimizing
  stabilizer parities for improved logical qubit memories},\ }\href
  {https://doi.org/10.1103/PhysRevLett.127.240501} {\bibfield  {journal}
  {\bibinfo  {journal} {Phys. Rev. Lett.}\ }\textbf {\bibinfo {volume} {127}},\
  \bibinfo {pages} {240501} (\bibinfo {year} {2021})}\BibitemShut {NoStop}%
\bibitem [{\citenamefont {Debroy}\ \emph {et~al.}(2018)\citenamefont {Debroy},
  \citenamefont {Li}, \citenamefont {Newman},\ and\ \citenamefont
  {Brown}}]{debroy2018}%
  \BibitemOpen
  \bibfield  {author} {\bibinfo {author} {\bibfnamefont {D.~M.}\ \bibnamefont
  {Debroy}}, \bibinfo {author} {\bibfnamefont {M.}~\bibnamefont {Li}}, \bibinfo
  {author} {\bibfnamefont {M.}~\bibnamefont {Newman}},\ and\ \bibinfo {author}
  {\bibfnamefont {K.~R.}\ \bibnamefont {Brown}},\ }\bibfield  {title} {\bibinfo
  {title} {Stabilizer slicing: Coherent error cancellations in low-density
  parity-check stabilizer codes},\ }\href
  {https://doi.org/10.1103/PhysRevLett.121.250502} {\bibfield  {journal}
  {\bibinfo  {journal} {Phys. Rev. Lett.}\ }\textbf {\bibinfo {volume} {121}},\
  \bibinfo {pages} {250502} (\bibinfo {year} {2018})}\BibitemShut {NoStop}%
\bibitem [{\citenamefont {Beale}\ \emph {et~al.}(2018)\citenamefont {Beale},
  \citenamefont {Wallman}, \citenamefont {Guti\'errez}, \citenamefont {Brown},\
  and\ \citenamefont {Laflamme}}]{beale2018}%
  \BibitemOpen
  \bibfield  {author} {\bibinfo {author} {\bibfnamefont {S.~J.}\ \bibnamefont
  {Beale}}, \bibinfo {author} {\bibfnamefont {J.~J.}\ \bibnamefont {Wallman}},
  \bibinfo {author} {\bibfnamefont {M.}~\bibnamefont {Guti\'errez}}, \bibinfo
  {author} {\bibfnamefont {K.~R.}\ \bibnamefont {Brown}},\ and\ \bibinfo
  {author} {\bibfnamefont {R.}~\bibnamefont {Laflamme}},\ }\bibfield  {title}
  {\bibinfo {title} {Quantum error correction decoheres noise},\ }\href
  {https://doi.org/10.1103/PhysRevLett.121.190501} {\bibfield  {journal}
  {\bibinfo  {journal} {Phys. Rev. Lett.}\ }\textbf {\bibinfo {volume} {121}},\
  \bibinfo {pages} {190501} (\bibinfo {year} {2018})}\BibitemShut {NoStop}%
\bibitem [{\citenamefont {Huang}\ \emph {et~al.}(2019)\citenamefont {Huang},
  \citenamefont {Doherty},\ and\ \citenamefont
  {Flammia}}]{huang2019performance}%
  \BibitemOpen
  \bibfield  {author} {\bibinfo {author} {\bibfnamefont {E.}~\bibnamefont
  {Huang}}, \bibinfo {author} {\bibfnamefont {A.~C.}\ \bibnamefont {Doherty}},\
  and\ \bibinfo {author} {\bibfnamefont {S.}~\bibnamefont {Flammia}},\
  }\bibfield  {title} {\bibinfo {title} {Performance of quantum error
  correction with coherent errors},\ }\href
  {https://doi.org/10.1103/PhysRevA.99.022313} {\bibfield  {journal} {\bibinfo
  {journal} {Phys. Rev. A}\ }\textbf {\bibinfo {volume} {99}},\ \bibinfo
  {pages} {022313} (\bibinfo {year} {2019})}\BibitemShut {NoStop}%
\bibitem [{\citenamefont {Iverson}\ and\ \citenamefont
  {Preskill}(2020)}]{iverson2020}%
  \BibitemOpen
  \bibfield  {author} {\bibinfo {author} {\bibfnamefont {J.~K.}\ \bibnamefont
  {Iverson}}\ and\ \bibinfo {author} {\bibfnamefont {J.}~\bibnamefont
  {Preskill}},\ }\bibfield  {title} {\bibinfo {title} {Coherence in logical
  quantum channels},\ }\href {https://doi.org/10.1088/1367-2630/ab8e5c}
  {\bibfield  {journal} {\bibinfo  {journal} {New Journal of Physics}\ }\textbf
  {\bibinfo {volume} {22}},\ \bibinfo {pages} {073066} (\bibinfo {year}
  {2020})}\BibitemShut {NoStop}%
\bibitem [{\citenamefont {Bruzewicz}\ \emph {et~al.}(2019)\citenamefont
  {Bruzewicz}, \citenamefont {Chiaverini}, \citenamefont {McConnell},\ and\
  \citenamefont {Sage}}]{Bruzewicz2019apr}%
  \BibitemOpen
  \bibfield  {author} {\bibinfo {author} {\bibfnamefont {C.~D.}\ \bibnamefont
  {Bruzewicz}}, \bibinfo {author} {\bibfnamefont {J.}~\bibnamefont
  {Chiaverini}}, \bibinfo {author} {\bibfnamefont {R.}~\bibnamefont
  {McConnell}},\ and\ \bibinfo {author} {\bibfnamefont {J.~M.}\ \bibnamefont
  {Sage}},\ }\bibfield  {title} {\bibinfo {title} {Trapped-ion quantum
  computing: Progress and challenges},\ }\href
  {https://doi.org/10.1063/1.5088164} {\bibfield  {journal} {\bibinfo
  {journal} {Applied Physics Reviews}\ }\textbf {\bibinfo {volume} {6}},\
  \bibinfo {pages} {021314} (\bibinfo {year} {2019})}\BibitemShut {NoStop}%
\bibitem [{\citenamefont {Kelly}\ \emph {et~al.}(2018)\citenamefont {Kelly},
  \citenamefont {O'Malley}, \citenamefont {Neeley}, \citenamefont {Neven},\
  and\ \citenamefont {Martinis}}]{kellyarXiv2018}%
  \BibitemOpen
  \bibfield  {author} {\bibinfo {author} {\bibfnamefont {J.}~\bibnamefont
  {Kelly}}, \bibinfo {author} {\bibfnamefont {P.}~\bibnamefont {O'Malley}},
  \bibinfo {author} {\bibfnamefont {M.}~\bibnamefont {Neeley}}, \bibinfo
  {author} {\bibfnamefont {H.}~\bibnamefont {Neven}},\ and\ \bibinfo {author}
  {\bibfnamefont {J.~M.}\ \bibnamefont {Martinis}},\ }\href@noop {} {\bibinfo
  {title} {Physical qubit calibration on a directed acyclic graph}} (\bibinfo
  {year} {2018}),\ \Eprint {https://arxiv.org/abs/1803.03226} {arXiv:1803.03226
  [quant-ph]} \BibitemShut {NoStop}%
\bibitem [{\citenamefont {Baum}\ \emph {et~al.}(2021)\citenamefont {Baum},
  \citenamefont {Amico}, \citenamefont {Howell}, \citenamefont {Hush},
  \citenamefont {Liuzzi}, \citenamefont {Mundada}, \citenamefont {Merkh},
  \citenamefont {Carvalho},\ and\ \citenamefont
  {Biercuk}}]{BaumPRXQuantum2021}%
  \BibitemOpen
  \bibfield  {author} {\bibinfo {author} {\bibfnamefont {Y.}~\bibnamefont
  {Baum}}, \bibinfo {author} {\bibfnamefont {M.}~\bibnamefont {Amico}},
  \bibinfo {author} {\bibfnamefont {S.}~\bibnamefont {Howell}}, \bibinfo
  {author} {\bibfnamefont {M.}~\bibnamefont {Hush}}, \bibinfo {author}
  {\bibfnamefont {M.}~\bibnamefont {Liuzzi}}, \bibinfo {author} {\bibfnamefont
  {P.}~\bibnamefont {Mundada}}, \bibinfo {author} {\bibfnamefont
  {T.}~\bibnamefont {Merkh}}, \bibinfo {author} {\bibfnamefont {A.~R.~R.}\
  \bibnamefont {Carvalho}},\ and\ \bibinfo {author} {\bibfnamefont {M.~J.}\
  \bibnamefont {Biercuk}},\ }\bibfield  {title} {\bibinfo {title} {Experimental
  deep reinforcement learning for error-robust gate-set design on a
  superconducting quantum computer},\ }\href
  {https://doi.org/10.1103/PRXQuantum.2.040324} {\bibfield  {journal} {\bibinfo
   {journal} {PRX Quantum}\ }\textbf {\bibinfo {volume} {2}},\ \bibinfo {pages}
  {040324} (\bibinfo {year} {2021})}\BibitemShut {NoStop}%
\bibitem [{\citenamefont {Nielsen}\ \emph {et~al.}(2020)\citenamefont
  {Nielsen}, \citenamefont {Rudinger}, \citenamefont {Proctor}, \citenamefont
  {Russo}, \citenamefont {Young},\ and\ \citenamefont
  {Blume-Kohout}}]{Nielsen2020}%
  \BibitemOpen
  \bibfield  {author} {\bibinfo {author} {\bibfnamefont {E.}~\bibnamefont
  {Nielsen}}, \bibinfo {author} {\bibfnamefont {K.}~\bibnamefont {Rudinger}},
  \bibinfo {author} {\bibfnamefont {T.}~\bibnamefont {Proctor}}, \bibinfo
  {author} {\bibfnamefont {A.}~\bibnamefont {Russo}}, \bibinfo {author}
  {\bibfnamefont {K.}~\bibnamefont {Young}},\ and\ \bibinfo {author}
  {\bibfnamefont {R.}~\bibnamefont {Blume-Kohout}},\ }\bibfield  {title}
  {\bibinfo {title} {Probing quantum processor performance with {pyGSTi}},\
  }\href {https://doi.org/10.1088/2058-9565/ab8aa4} {\bibfield  {journal}
  {\bibinfo  {journal} {Quantum Science and Technology}\ }\textbf {\bibinfo
  {volume} {5}},\ \bibinfo {pages} {044002} (\bibinfo {year}
  {2020})}\BibitemShut {NoStop}%
\bibitem [{\citenamefont {Whitfield}\ \emph {et~al.}(2011)\citenamefont
  {Whitfield}, \citenamefont {Biamonte},\ and\ \citenamefont
  {Aspuru-Guzik}}]{whitfield2011simulation}%
  \BibitemOpen
  \bibfield  {author} {\bibinfo {author} {\bibfnamefont {J.~D.}\ \bibnamefont
  {Whitfield}}, \bibinfo {author} {\bibfnamefont {J.}~\bibnamefont
  {Biamonte}},\ and\ \bibinfo {author} {\bibfnamefont {A.}~\bibnamefont
  {Aspuru-Guzik}},\ }\bibfield  {title} {\bibinfo {title} {Simulation of
  electronic structure hamiltonians using quantum computers},\ }\href@noop {}
  {\bibfield  {journal} {\bibinfo  {journal} {Mol. Phys.}\ }\textbf {\bibinfo
  {volume} {109}},\ \bibinfo {pages} {735} (\bibinfo {year}
  {2011})}\BibitemShut {NoStop}%
\bibitem [{\citenamefont {Khaneja}\ \emph {et~al.}(2001)\citenamefont
  {Khaneja}, \citenamefont {Brockett},\ and\ \citenamefont
  {Glaser}}]{KhanejaPhysRevA2001}%
  \BibitemOpen
  \bibfield  {author} {\bibinfo {author} {\bibfnamefont {N.}~\bibnamefont
  {Khaneja}}, \bibinfo {author} {\bibfnamefont {R.}~\bibnamefont {Brockett}},\
  and\ \bibinfo {author} {\bibfnamefont {S.~J.}\ \bibnamefont {Glaser}},\
  }\bibfield  {title} {\bibinfo {title} {Time optimal control in spin
  systems},\ }\href {https://doi.org/10.1103/PhysRevA.63.032308} {\bibfield
  {journal} {\bibinfo  {journal} {Phys. Rev. A}\ }\textbf {\bibinfo {volume}
  {63}},\ \bibinfo {pages} {032308} (\bibinfo {year} {2001})}\BibitemShut
  {NoStop}%
\bibitem [{\citenamefont {Shi}\ \emph {et~al.}(2019)\citenamefont {Shi},
  \citenamefont {Leung}, \citenamefont {Gokhale}, \citenamefont {Rossi},
  \citenamefont {Schuster}, \citenamefont {Hoffmann},\ and\ \citenamefont
  {Chong}}]{ShiASPLOS2019}%
  \BibitemOpen
  \bibfield  {author} {\bibinfo {author} {\bibfnamefont {Y.}~\bibnamefont
  {Shi}}, \bibinfo {author} {\bibfnamefont {N.}~\bibnamefont {Leung}}, \bibinfo
  {author} {\bibfnamefont {P.}~\bibnamefont {Gokhale}}, \bibinfo {author}
  {\bibfnamefont {Z.}~\bibnamefont {Rossi}}, \bibinfo {author} {\bibfnamefont
  {D.~I.}\ \bibnamefont {Schuster}}, \bibinfo {author} {\bibfnamefont
  {H.}~\bibnamefont {Hoffmann}},\ and\ \bibinfo {author} {\bibfnamefont
  {F.~T.~C.}\ \bibnamefont {Chong}},\ }\bibfield  {title} {\bibinfo {title}
  {Optimized compilation of aggregated instructions for realistic quantum
  computers},\ }\href@noop {} {\bibfield  {journal} {\bibinfo  {journal}
  {Proceedings of the Twenty-Fourth International Conference on Architectural
  Support for Programming Languages and Operating Systems}\ ,\ \bibinfo {pages}
  {1031}} (\bibinfo {year} {2019})}\BibitemShut {NoStop}%
\bibitem [{\citenamefont {Murali}\ \emph {et~al.}(2019)\citenamefont {Murali},
  \citenamefont {Baker}, \citenamefont {Javadi-Abhari}, \citenamefont {Chong},\
  and\ \citenamefont {Martonosi}}]{MuraliASPLOS2019}%
  \BibitemOpen
  \bibfield  {author} {\bibinfo {author} {\bibfnamefont {P.}~\bibnamefont
  {Murali}}, \bibinfo {author} {\bibfnamefont {J.~M.}\ \bibnamefont {Baker}},
  \bibinfo {author} {\bibfnamefont {A.}~\bibnamefont {Javadi-Abhari}}, \bibinfo
  {author} {\bibfnamefont {F.~T.}\ \bibnamefont {Chong}},\ and\ \bibinfo
  {author} {\bibfnamefont {M.}~\bibnamefont {Martonosi}},\ }\bibfield  {title}
  {\bibinfo {title} {Noise-adaptive compiler mappings for noisy
  intermediate-scale quantum computers},\ }\href@noop {} {\bibfield  {journal}
  {\bibinfo  {journal} {Proceedings of the Twenty-Fourth International
  Conference on Architectural Support for Programming Languages and Operating
  Systems}\ ,\ \bibinfo {pages} {1015}} (\bibinfo {year} {2019})}\BibitemShut
  {NoStop}%
\bibitem [{\citenamefont {Gokhale}\ \emph {et~al.}(2019)\citenamefont
  {Gokhale}, \citenamefont {Ding}, \citenamefont {Propson}, \citenamefont
  {Winkler}, \citenamefont {Leung}, \citenamefont {Shi}, \citenamefont
  {Schuster}, \citenamefont {Hoffmann},\ and\ \citenamefont
  {Chong}}]{GokhaleMicro2019}%
  \BibitemOpen
  \bibfield  {author} {\bibinfo {author} {\bibfnamefont {P.}~\bibnamefont
  {Gokhale}}, \bibinfo {author} {\bibfnamefont {Y.}~\bibnamefont {Ding}},
  \bibinfo {author} {\bibfnamefont {T.}~\bibnamefont {Propson}}, \bibinfo
  {author} {\bibfnamefont {C.}~\bibnamefont {Winkler}}, \bibinfo {author}
  {\bibfnamefont {N.}~\bibnamefont {Leung}}, \bibinfo {author} {\bibfnamefont
  {Y.}~\bibnamefont {Shi}}, \bibinfo {author} {\bibfnamefont {D.~I.}\
  \bibnamefont {Schuster}}, \bibinfo {author} {\bibfnamefont {H.}~\bibnamefont
  {Hoffmann}},\ and\ \bibinfo {author} {\bibfnamefont {F.~T.}\ \bibnamefont
  {Chong}},\ }\bibfield  {title} {\bibinfo {title} {Partial compilation of
  variational algorithms for noisy intermediate-scale quantum machines},\ }in\
  \href {https://doi.org/10.1145/3352460.3358313} {\emph {\bibinfo {booktitle}
  {Proceedings of the 52nd Annual IEEE/ACM International Symposium on
  Microarchitecture}}},\ \bibinfo {series and number} {MICRO '52}\ (\bibinfo
  {publisher} {Association for Computing Machinery},\ \bibinfo {address} {New
  York, NY, USA},\ \bibinfo {year} {2019})\ p.\ \bibinfo {pages}
  {266–278}\BibitemShut {NoStop}%
\bibitem [{\citenamefont {Magann}\ \emph {et~al.}(2021)\citenamefont {Magann},
  \citenamefont {Arenz}, \citenamefont {Grace}, \citenamefont {Ho},
  \citenamefont {Kosut}, \citenamefont {McClean}, \citenamefont {Rabitz},\ and\
  \citenamefont {Sarovar}}]{MagannPRXQuantum2021}%
  \BibitemOpen
  \bibfield  {author} {\bibinfo {author} {\bibfnamefont {A.~B.}\ \bibnamefont
  {Magann}}, \bibinfo {author} {\bibfnamefont {C.}~\bibnamefont {Arenz}},
  \bibinfo {author} {\bibfnamefont {M.~D.}\ \bibnamefont {Grace}}, \bibinfo
  {author} {\bibfnamefont {T.-S.}\ \bibnamefont {Ho}}, \bibinfo {author}
  {\bibfnamefont {R.~L.}\ \bibnamefont {Kosut}}, \bibinfo {author}
  {\bibfnamefont {J.~R.}\ \bibnamefont {McClean}}, \bibinfo {author}
  {\bibfnamefont {H.~A.}\ \bibnamefont {Rabitz}},\ and\ \bibinfo {author}
  {\bibfnamefont {M.}~\bibnamefont {Sarovar}},\ }\bibfield  {title} {\bibinfo
  {title} {From pulses to circuits and back again: A quantum optimal control
  perspective on variational quantum algorithms},\ }\href
  {https://doi.org/10.1103/PRXQuantum.2.010101} {\bibfield  {journal} {\bibinfo
   {journal} {PRX Quantum}\ }\textbf {\bibinfo {volume} {2}},\ \bibinfo {pages}
  {010101} (\bibinfo {year} {2021})}\BibitemShut {NoStop}%
\bibitem [{\citenamefont {Meitei}\ \emph {et~al.}(2021)\citenamefont {Meitei},
  \citenamefont {Gard}, \citenamefont {Barron}, \citenamefont {Pappas},
  \citenamefont {Economou}, \citenamefont {Barnes},\ and\ \citenamefont
  {Mayhall}}]{meiteiArXiv2021}%
  \BibitemOpen
  \bibfield  {author} {\bibinfo {author} {\bibfnamefont {O.~R.}\ \bibnamefont
  {Meitei}}, \bibinfo {author} {\bibfnamefont {B.~T.}\ \bibnamefont {Gard}},
  \bibinfo {author} {\bibfnamefont {G.~S.}\ \bibnamefont {Barron}}, \bibinfo
  {author} {\bibfnamefont {D.~P.}\ \bibnamefont {Pappas}}, \bibinfo {author}
  {\bibfnamefont {S.~E.}\ \bibnamefont {Economou}}, \bibinfo {author}
  {\bibfnamefont {E.}~\bibnamefont {Barnes}},\ and\ \bibinfo {author}
  {\bibfnamefont {N.~J.}\ \bibnamefont {Mayhall}},\ }\bibfield  {title}
  {\bibinfo {title} {Gate-free state preparation for fast variational quantum
  eigensolver simulations},\ }\href@noop {} {\bibfield  {journal} {\bibinfo
  {journal} {npj Quantum Information}\ }\textbf {\bibinfo {volume} {7}},\
  \bibinfo {pages} {1} (\bibinfo {year} {2021})}\BibitemShut {NoStop}%
\bibitem [{\citenamefont {Maslov}(2017)}]{maslov2017}%
  \BibitemOpen
  \bibfield  {author} {\bibinfo {author} {\bibfnamefont {D.}~\bibnamefont
  {Maslov}},\ }\bibfield  {title} {\bibinfo {title} {Basic circuit compilation
  techniques for an ion-trap quantum machine},\ }\href
  {https://doi.org/10.1088/1367-2630/aa5e47} {\bibfield  {journal} {\bibinfo
  {journal} {New Journal of Physics}\ }\textbf {\bibinfo {volume} {19}},\
  \bibinfo {pages} {023035} (\bibinfo {year} {2017})}\BibitemShut {NoStop}%
\bibitem [{\citenamefont {Madzik}\ \emph {et~al.}(2021)\citenamefont {Madzik},
  \citenamefont {Laucht}, \citenamefont {Hudson}, \citenamefont {Jakob},
  \citenamefont {Johnson}, \citenamefont {Jamieson}, \citenamefont {Itoh},
  \citenamefont {Dzurak},\ and\ \citenamefont {Morello}}]{MadzikNatComm2021}%
  \BibitemOpen
  \bibfield  {author} {\bibinfo {author} {\bibfnamefont {M.~T.}\ \bibnamefont
  {Madzik}}, \bibinfo {author} {\bibfnamefont {A.}~\bibnamefont {Laucht}},
  \bibinfo {author} {\bibfnamefont {F.~E.}\ \bibnamefont {Hudson}}, \bibinfo
  {author} {\bibfnamefont {A.~M.}\ \bibnamefont {Jakob}}, \bibinfo {author}
  {\bibfnamefont {B.~C.}\ \bibnamefont {Johnson}}, \bibinfo {author}
  {\bibfnamefont {D.~N.}\ \bibnamefont {Jamieson}}, \bibinfo {author}
  {\bibfnamefont {K.~M.}\ \bibnamefont {Itoh}}, \bibinfo {author}
  {\bibfnamefont {A.~S.}\ \bibnamefont {Dzurak}},\ and\ \bibinfo {author}
  {\bibfnamefont {A.}~\bibnamefont {Morello}},\ }\bibfield  {title} {\bibinfo
  {title} {Conditional quantum operation of two exchange-coupled single-donor
  spin qubits in a mos-compatible silicon device},\ }\href@noop {} {\bibfield
  {journal} {\bibinfo  {journal} {Nature communications}\ }\textbf {\bibinfo
  {volume} {12}},\ \bibinfo {pages} {1} (\bibinfo {year} {2021})}\BibitemShut
  {NoStop}%
\bibitem [{\citenamefont {Barkoutsos}\ \emph {et~al.}(2018)\citenamefont
  {Barkoutsos}, \citenamefont {Gonthier}, \citenamefont {Sokolov},
  \citenamefont {Moll}, \citenamefont {Salis}, \citenamefont {Fuhrer},
  \citenamefont {Ganzhorn}, \citenamefont {Egger}, \citenamefont {Troyer},
  \citenamefont {Mezzacapo}, \citenamefont {Filipp},\ and\ \citenamefont
  {Tavernelli}}]{barkoutsos2018}%
  \BibitemOpen
  \bibfield  {author} {\bibinfo {author} {\bibfnamefont {P.~K.}\ \bibnamefont
  {Barkoutsos}}, \bibinfo {author} {\bibfnamefont {J.~F.}\ \bibnamefont
  {Gonthier}}, \bibinfo {author} {\bibfnamefont {I.}~\bibnamefont {Sokolov}},
  \bibinfo {author} {\bibfnamefont {N.}~\bibnamefont {Moll}}, \bibinfo {author}
  {\bibfnamefont {G.}~\bibnamefont {Salis}}, \bibinfo {author} {\bibfnamefont
  {A.}~\bibnamefont {Fuhrer}}, \bibinfo {author} {\bibfnamefont
  {M.}~\bibnamefont {Ganzhorn}}, \bibinfo {author} {\bibfnamefont {D.~J.}\
  \bibnamefont {Egger}}, \bibinfo {author} {\bibfnamefont {M.}~\bibnamefont
  {Troyer}}, \bibinfo {author} {\bibfnamefont {A.}~\bibnamefont {Mezzacapo}},
  \bibinfo {author} {\bibfnamefont {S.}~\bibnamefont {Filipp}},\ and\ \bibinfo
  {author} {\bibfnamefont {I.}~\bibnamefont {Tavernelli}},\ }\bibfield  {title}
  {\bibinfo {title} {Quantum algorithms for electronic structure calculations:
  Particle-hole hamiltonian and optimized wave-function expansions},\ }\href
  {https://doi.org/10.1103/PhysRevA.98.022322} {\bibfield  {journal} {\bibinfo
  {journal} {Phys. Rev. A}\ }\textbf {\bibinfo {volume} {98}},\ \bibinfo
  {pages} {022322} (\bibinfo {year} {2018})}\BibitemShut {NoStop}%
\bibitem [{\citenamefont {Barnes}\ \emph {et~al.}(2017)\citenamefont {Barnes},
  \citenamefont {Trout}, \citenamefont {Lucarelli},\ and\ \citenamefont
  {Clader}}]{BarnesPRA2017}%
  \BibitemOpen
  \bibfield  {author} {\bibinfo {author} {\bibfnamefont {J.~P.}\ \bibnamefont
  {Barnes}}, \bibinfo {author} {\bibfnamefont {C.~J.}\ \bibnamefont {Trout}},
  \bibinfo {author} {\bibfnamefont {D.}~\bibnamefont {Lucarelli}},\ and\
  \bibinfo {author} {\bibfnamefont {B.~D.}\ \bibnamefont {Clader}},\ }\bibfield
   {title} {\bibinfo {title} {Quantum error-correction failure distributions:
  Comparison of coherent and stochastic error models},\ }\href
  {https://doi.org/10.1103/PhysRevA.95.062338} {\bibfield  {journal} {\bibinfo
  {journal} {Phys. Rev. A}\ }\textbf {\bibinfo {volume} {95}},\ \bibinfo
  {pages} {062338} (\bibinfo {year} {2017})}\BibitemShut {NoStop}%
\bibitem [{\citenamefont {S\o{}rensen}\ and\ \citenamefont
  {M\o{}lmer}(1999)}]{sorensen1999}%
  \BibitemOpen
  \bibfield  {author} {\bibinfo {author} {\bibfnamefont {A.}~\bibnamefont
  {S\o{}rensen}}\ and\ \bibinfo {author} {\bibfnamefont {K.}~\bibnamefont
  {M\o{}lmer}},\ }\bibfield  {title} {\bibinfo {title} {Quantum computation
  with ions in thermal motion},\ }\href
  {https://doi.org/10.1103/PhysRevLett.82.1971} {\bibfield  {journal} {\bibinfo
   {journal} {Phys. Rev. Lett.}\ }\textbf {\bibinfo {volume} {82}},\ \bibinfo
  {pages} {1971} (\bibinfo {year} {1999})}\BibitemShut {NoStop}%
\bibitem [{\citenamefont {McKeeman}(1965)}]{McKeeman65}%
  \BibitemOpen
  \bibfield  {author} {\bibinfo {author} {\bibfnamefont {W.~M.}\ \bibnamefont
  {McKeeman}},\ }\bibfield  {title} {\bibinfo {title} {Peephole optimization},\
  }\href {https://doi.org/10.1145/364995.365000} {\bibfield  {journal}
  {\bibinfo  {journal} {Commun. ACM}\ }\textbf {\bibinfo {volume} {8}},\
  \bibinfo {pages} {443–444} (\bibinfo {year} {1965})}\BibitemShut {NoStop}%
\bibitem [{\citenamefont {Wang}\ \emph {et~al.}(2020)\citenamefont {Wang},
  \citenamefont {Crain}, \citenamefont {Fang}, \citenamefont {Zhang},
  \citenamefont {Huang}, \citenamefont {Liang}, \citenamefont {Leung},
  \citenamefont {Brown},\ and\ \citenamefont {Kim}}]{wang2020}%
  \BibitemOpen
  \bibfield  {author} {\bibinfo {author} {\bibfnamefont {Y.}~\bibnamefont
  {Wang}}, \bibinfo {author} {\bibfnamefont {S.}~\bibnamefont {Crain}},
  \bibinfo {author} {\bibfnamefont {C.}~\bibnamefont {Fang}}, \bibinfo {author}
  {\bibfnamefont {B.}~\bibnamefont {Zhang}}, \bibinfo {author} {\bibfnamefont
  {S.}~\bibnamefont {Huang}}, \bibinfo {author} {\bibfnamefont
  {Q.}~\bibnamefont {Liang}}, \bibinfo {author} {\bibfnamefont {P.~H.}\
  \bibnamefont {Leung}}, \bibinfo {author} {\bibfnamefont {K.~R.}\ \bibnamefont
  {Brown}},\ and\ \bibinfo {author} {\bibfnamefont {J.}~\bibnamefont {Kim}},\
  }\bibfield  {title} {\bibinfo {title} {High-fidelity two-qubit gates using a
  microelectromechanical-system-based beam steering system for individual qubit
  addressing},\ }\href {https://doi.org/10.1103/PhysRevLett.125.150505}
  {\bibfield  {journal} {\bibinfo  {journal} {Phys. Rev. Lett.}\ }\textbf
  {\bibinfo {volume} {125}},\ \bibinfo {pages} {150505} (\bibinfo {year}
  {2020})}\BibitemShut {NoStop}%
\bibitem [{\citenamefont {Noek}\ \emph {et~al.}(2013)\citenamefont {Noek},
  \citenamefont {Vrijsen}, \citenamefont {Gaultney}, \citenamefont {Mount},
  \citenamefont {Kim}, \citenamefont {Maunz},\ and\ \citenamefont
  {Kim}}]{Noek2013}%
  \BibitemOpen
  \bibfield  {author} {\bibinfo {author} {\bibfnamefont {R.}~\bibnamefont
  {Noek}}, \bibinfo {author} {\bibfnamefont {G.}~\bibnamefont {Vrijsen}},
  \bibinfo {author} {\bibfnamefont {D.}~\bibnamefont {Gaultney}}, \bibinfo
  {author} {\bibfnamefont {E.}~\bibnamefont {Mount}}, \bibinfo {author}
  {\bibfnamefont {T.}~\bibnamefont {Kim}}, \bibinfo {author} {\bibfnamefont
  {P.}~\bibnamefont {Maunz}},\ and\ \bibinfo {author} {\bibfnamefont
  {J.}~\bibnamefont {Kim}},\ }\bibfield  {title} {\bibinfo {title} {High speed,
  high fidelity detection of an atomic hyperfine qubit},\ }\href
  {https://doi.org/10.1364/OL.38.004735} {\bibfield  {journal} {\bibinfo
  {journal} {Opt. Lett.}\ }\textbf {\bibinfo {volume} {38}},\ \bibinfo {pages}
  {4735} (\bibinfo {year} {2013})}\BibitemShut {NoStop}%
\bibitem [{\citenamefont {Myerson}\ \emph {et~al.}(2008)\citenamefont
  {Myerson}, \citenamefont {Szwer}, \citenamefont {Webster}, \citenamefont
  {Allcock}, \citenamefont {Curtis}, \citenamefont {Imreh}, \citenamefont
  {Sherman}, \citenamefont {Stacey}, \citenamefont {Steane},\ and\
  \citenamefont {Lucas}}]{Myerson2008}%
  \BibitemOpen
  \bibfield  {author} {\bibinfo {author} {\bibfnamefont {A.~H.}\ \bibnamefont
  {Myerson}}, \bibinfo {author} {\bibfnamefont {D.~J.}\ \bibnamefont {Szwer}},
  \bibinfo {author} {\bibfnamefont {S.~C.}\ \bibnamefont {Webster}}, \bibinfo
  {author} {\bibfnamefont {D.~T.~C.}\ \bibnamefont {Allcock}}, \bibinfo
  {author} {\bibfnamefont {M.~J.}\ \bibnamefont {Curtis}}, \bibinfo {author}
  {\bibfnamefont {G.}~\bibnamefont {Imreh}}, \bibinfo {author} {\bibfnamefont
  {J.~A.}\ \bibnamefont {Sherman}}, \bibinfo {author} {\bibfnamefont {D.~N.}\
  \bibnamefont {Stacey}}, \bibinfo {author} {\bibfnamefont {A.~M.}\
  \bibnamefont {Steane}},\ and\ \bibinfo {author} {\bibfnamefont {D.~M.}\
  \bibnamefont {Lucas}},\ }\bibfield  {title} {\bibinfo {title} {High-fidelity
  readout of trapped-ion qubits},\ }\href
  {https://doi.org/10.1103/PhysRevLett.100.200502} {\bibfield  {journal}
  {\bibinfo  {journal} {Phys. Rev. Lett.}\ }\textbf {\bibinfo {volume} {100}},\
  \bibinfo {pages} {200502} (\bibinfo {year} {2008})}\BibitemShut {NoStop}%
\bibitem [{\citenamefont {Crain}\ \emph {et~al.}(2019)\citenamefont {Crain},
  \citenamefont {Cahall}, \citenamefont {Vrijsen}, \citenamefont {Wollman},
  \citenamefont {Shaw}, \citenamefont {Verma}, \citenamefont {Nam},\ and\
  \citenamefont {Kim}}]{crain2019high}%
  \BibitemOpen
  \bibfield  {author} {\bibinfo {author} {\bibfnamefont {S.}~\bibnamefont
  {Crain}}, \bibinfo {author} {\bibfnamefont {C.}~\bibnamefont {Cahall}},
  \bibinfo {author} {\bibfnamefont {G.}~\bibnamefont {Vrijsen}}, \bibinfo
  {author} {\bibfnamefont {E.~E.}\ \bibnamefont {Wollman}}, \bibinfo {author}
  {\bibfnamefont {M.~D.}\ \bibnamefont {Shaw}}, \bibinfo {author}
  {\bibfnamefont {V.~B.}\ \bibnamefont {Verma}}, \bibinfo {author}
  {\bibfnamefont {S.~W.}\ \bibnamefont {Nam}},\ and\ \bibinfo {author}
  {\bibfnamefont {J.}~\bibnamefont {Kim}},\ }\bibfield  {title} {\bibinfo
  {title} {High-speed low-crosstalk detection of a 171 yb+ qubit using
  superconducting nanowire single photon detectors},\ }\href@noop {} {\bibfield
   {journal} {\bibinfo  {journal} {Communications Physics}\ }\textbf {\bibinfo
  {volume} {2}},\ \bibinfo {pages} {1} (\bibinfo {year} {2019})}\BibitemShut
  {NoStop}%
\bibitem [{\citenamefont {Wineland}\ \emph {et~al.}(1998)\citenamefont
  {Wineland}, \citenamefont {Monroe}, \citenamefont {Itano}, \citenamefont
  {Leibfried}, \citenamefont {King},\ and\ \citenamefont
  {Meekhof}}]{wineland1998experimental}%
  \BibitemOpen
  \bibfield  {author} {\bibinfo {author} {\bibfnamefont {D.~J.}\ \bibnamefont
  {Wineland}}, \bibinfo {author} {\bibfnamefont {C.}~\bibnamefont {Monroe}},
  \bibinfo {author} {\bibfnamefont {W.~M.}\ \bibnamefont {Itano}}, \bibinfo
  {author} {\bibfnamefont {D.}~\bibnamefont {Leibfried}}, \bibinfo {author}
  {\bibfnamefont {B.~E.}\ \bibnamefont {King}},\ and\ \bibinfo {author}
  {\bibfnamefont {D.~M.}\ \bibnamefont {Meekhof}},\ }\bibfield  {title}
  {\bibinfo {title} {Experimental issues in coherent quantum-state manipulation
  of trapped atomic ions},\ }\href@noop {} {\bibfield  {journal} {\bibinfo
  {journal} {J. Res. Natl. Inst. Stand. Technol.}\ }\textbf {\bibinfo {volume}
  {103}},\ \bibinfo {pages} {259} (\bibinfo {year} {1998})}\BibitemShut
  {NoStop}%
\bibitem [{\citenamefont {Mount}\ \emph {et~al.}(2013)\citenamefont {Mount},
  \citenamefont {Baek}, \citenamefont {Blain}, \citenamefont {Stick},
  \citenamefont {Gaultney}, \citenamefont {Crain}, \citenamefont {Noek},
  \citenamefont {Kim}, \citenamefont {Maunz},\ and\ \citenamefont
  {Kim}}]{Mount2013}%
  \BibitemOpen
  \bibfield  {author} {\bibinfo {author} {\bibfnamefont {E.}~\bibnamefont
  {Mount}}, \bibinfo {author} {\bibfnamefont {S.-Y.}\ \bibnamefont {Baek}},
  \bibinfo {author} {\bibfnamefont {M.}~\bibnamefont {Blain}}, \bibinfo
  {author} {\bibfnamefont {D.}~\bibnamefont {Stick}}, \bibinfo {author}
  {\bibfnamefont {D.}~\bibnamefont {Gaultney}}, \bibinfo {author}
  {\bibfnamefont {S.}~\bibnamefont {Crain}}, \bibinfo {author} {\bibfnamefont
  {R.}~\bibnamefont {Noek}}, \bibinfo {author} {\bibfnamefont {T.}~\bibnamefont
  {Kim}}, \bibinfo {author} {\bibfnamefont {P.}~\bibnamefont {Maunz}},\ and\
  \bibinfo {author} {\bibfnamefont {J.}~\bibnamefont {Kim}},\ }\bibfield
  {title} {\bibinfo {title} {Single qubit manipulation in a microfabricated
  surface electrode ion trap},\ }\href
  {https://doi.org/10.1088/1367-2630/15/9/093018} {\bibfield  {journal}
  {\bibinfo  {journal} {New Journal of Physics}\ }\textbf {\bibinfo {volume}
  {15}},\ \bibinfo {pages} {093018} (\bibinfo {year} {2013})}\BibitemShut
  {NoStop}%
\bibitem [{\citenamefont {Inlek}\ \emph {et~al.}(2014)\citenamefont {Inlek},
  \citenamefont {Vittorini}, \citenamefont {Hucul}, \citenamefont {Crocker},\
  and\ \citenamefont {Monroe}}]{inlek2014}%
  \BibitemOpen
  \bibfield  {author} {\bibinfo {author} {\bibfnamefont {I.~V.}\ \bibnamefont
  {Inlek}}, \bibinfo {author} {\bibfnamefont {G.}~\bibnamefont {Vittorini}},
  \bibinfo {author} {\bibfnamefont {D.}~\bibnamefont {Hucul}}, \bibinfo
  {author} {\bibfnamefont {C.}~\bibnamefont {Crocker}},\ and\ \bibinfo {author}
  {\bibfnamefont {C.}~\bibnamefont {Monroe}},\ }\bibfield  {title} {\bibinfo
  {title} {Quantum gates with phase stability over space and time},\ }\href
  {https://doi.org/10.1103/PhysRevA.90.042316} {\bibfield  {journal} {\bibinfo
  {journal} {Phys. Rev. A}\ }\textbf {\bibinfo {volume} {90}},\ \bibinfo
  {pages} {042316} (\bibinfo {year} {2014})}\BibitemShut {NoStop}%
\bibitem [{\citenamefont {Crain}\ \emph {et~al.}(2014)\citenamefont {Crain},
  \citenamefont {Mount}, \citenamefont {Baek},\ and\ \citenamefont
  {Kim}}]{crain2014individual}%
  \BibitemOpen
  \bibfield  {author} {\bibinfo {author} {\bibfnamefont {S.}~\bibnamefont
  {Crain}}, \bibinfo {author} {\bibfnamefont {E.}~\bibnamefont {Mount}},
  \bibinfo {author} {\bibfnamefont {S.}~\bibnamefont {Baek}},\ and\ \bibinfo
  {author} {\bibfnamefont {J.}~\bibnamefont {Kim}},\ }\bibfield  {title}
  {\bibinfo {title} {Individual addressing of trapped 171yb+ ion qubits using a
  microelectromechanical systems-based beam steering system},\ }\href@noop {}
  {\bibfield  {journal} {\bibinfo  {journal} {Appl. Phys. Lett.}\ }\textbf
  {\bibinfo {volume} {105}},\ \bibinfo {pages} {181115} (\bibinfo {year}
  {2014})}\BibitemShut {NoStop}%
\bibitem [{\citenamefont {Clark}\ \emph {et~al.}(2021)\citenamefont {Clark},
  \citenamefont {Lobser}, \citenamefont {Revelle}, \citenamefont {Yale},
  \citenamefont {Bossert}, \citenamefont {Burch}, \citenamefont {Chow},
  \citenamefont {Hogle}, \citenamefont {Ivory}, \citenamefont {Pehr} \emph
  {et~al.}}]{Clark2021}%
  \BibitemOpen
  \bibfield  {author} {\bibinfo {author} {\bibfnamefont {S.~M.}\ \bibnamefont
  {Clark}}, \bibinfo {author} {\bibfnamefont {D.}~\bibnamefont {Lobser}},
  \bibinfo {author} {\bibfnamefont {M.~C.}\ \bibnamefont {Revelle}}, \bibinfo
  {author} {\bibfnamefont {C.~G.}\ \bibnamefont {Yale}}, \bibinfo {author}
  {\bibfnamefont {D.}~\bibnamefont {Bossert}}, \bibinfo {author} {\bibfnamefont
  {A.~D.}\ \bibnamefont {Burch}}, \bibinfo {author} {\bibfnamefont {M.~N.}\
  \bibnamefont {Chow}}, \bibinfo {author} {\bibfnamefont {C.~W.}\ \bibnamefont
  {Hogle}}, \bibinfo {author} {\bibfnamefont {M.}~\bibnamefont {Ivory}},
  \bibinfo {author} {\bibfnamefont {J.}~\bibnamefont {Pehr}}, \emph {et~al.},\
  }\bibfield  {title} {\bibinfo {title} {Engineering the quantum scientific
  computing open user testbed},\ }\href@noop {} {\bibfield  {journal} {\bibinfo
   {journal} {IEEE Transactions on Quantum Engineering}\ }\textbf {\bibinfo
  {volume} {2}},\ \bibinfo {pages} {1} (\bibinfo {year} {2021})}\BibitemShut
  {NoStop}%
\bibitem [{\citenamefont {S\o{}rensen}\ and\ \citenamefont
  {M\o{}lmer}(2000)}]{Sorensen2000}%
  \BibitemOpen
  \bibfield  {author} {\bibinfo {author} {\bibfnamefont {A.}~\bibnamefont
  {S\o{}rensen}}\ and\ \bibinfo {author} {\bibfnamefont {K.}~\bibnamefont
  {M\o{}lmer}},\ }\bibfield  {title} {\bibinfo {title} {Entanglement and
  quantum computation with ions in thermal motion},\ }\href
  {https://doi.org/10.1103/PhysRevA.62.022311} {\bibfield  {journal} {\bibinfo
  {journal} {Phys. Rev. A}\ }\textbf {\bibinfo {volume} {62}},\ \bibinfo
  {pages} {022311} (\bibinfo {year} {2000})}\BibitemShut {NoStop}%
\bibitem [{\citenamefont {Leung}\ \emph {et~al.}(2018)\citenamefont {Leung},
  \citenamefont {Landsman}, \citenamefont {Figgatt}, \citenamefont {Linke},
  \citenamefont {Monroe},\ and\ \citenamefont {Brown}}]{Leung2018}%
  \BibitemOpen
  \bibfield  {author} {\bibinfo {author} {\bibfnamefont {P.~H.}\ \bibnamefont
  {Leung}}, \bibinfo {author} {\bibfnamefont {K.~A.}\ \bibnamefont {Landsman}},
  \bibinfo {author} {\bibfnamefont {C.}~\bibnamefont {Figgatt}}, \bibinfo
  {author} {\bibfnamefont {N.~M.}\ \bibnamefont {Linke}}, \bibinfo {author}
  {\bibfnamefont {C.}~\bibnamefont {Monroe}},\ and\ \bibinfo {author}
  {\bibfnamefont {K.~R.}\ \bibnamefont {Brown}},\ }\bibfield  {title} {\bibinfo
  {title} {Robust 2-qubit gates in a linear ion crystal using a
  frequency-modulated driving force},\ }\href
  {https://doi.org/10.1103/PhysRevLett.120.020501} {\bibfield  {journal}
  {\bibinfo  {journal} {Phys. Rev. Lett.}\ }\textbf {\bibinfo {volume} {120}},\
  \bibinfo {pages} {020501} (\bibinfo {year} {2018})}\BibitemShut {NoStop}%
\bibitem [{\citenamefont {Landsman}\ \emph {et~al.}(2019)\citenamefont
  {Landsman}, \citenamefont {Wu}, \citenamefont {Leung}, \citenamefont {Zhu},
  \citenamefont {Linke}, \citenamefont {Brown}, \citenamefont {Duan},\ and\
  \citenamefont {Monroe}}]{Landsman2019}%
  \BibitemOpen
  \bibfield  {author} {\bibinfo {author} {\bibfnamefont {K.~A.}\ \bibnamefont
  {Landsman}}, \bibinfo {author} {\bibfnamefont {Y.}~\bibnamefont {Wu}},
  \bibinfo {author} {\bibfnamefont {P.~H.}\ \bibnamefont {Leung}}, \bibinfo
  {author} {\bibfnamefont {D.}~\bibnamefont {Zhu}}, \bibinfo {author}
  {\bibfnamefont {N.~M.}\ \bibnamefont {Linke}}, \bibinfo {author}
  {\bibfnamefont {K.~R.}\ \bibnamefont {Brown}}, \bibinfo {author}
  {\bibfnamefont {L.}~\bibnamefont {Duan}},\ and\ \bibinfo {author}
  {\bibfnamefont {C.}~\bibnamefont {Monroe}},\ }\bibfield  {title} {\bibinfo
  {title} {Two-qubit entangling gates within arbitrarily long chains of trapped
  ions},\ }\href {https://doi.org/10.1103/PhysRevA.100.022332} {\bibfield
  {journal} {\bibinfo  {journal} {Phys. Rev. A}\ }\textbf {\bibinfo {volume}
  {100}},\ \bibinfo {pages} {022332} (\bibinfo {year} {2019})}\BibitemShut
  {NoStop}%
\bibitem [{\citenamefont {Kang}\ \emph {et~al.}(2021)\citenamefont {Kang},
  \citenamefont {Liang}, \citenamefont {Zhang}, \citenamefont {Huang},
  \citenamefont {Wang}, \citenamefont {Fang}, \citenamefont {Kim},\ and\
  \citenamefont {Brown}}]{kang2021batch}%
  \BibitemOpen
  \bibfield  {author} {\bibinfo {author} {\bibfnamefont {M.}~\bibnamefont
  {Kang}}, \bibinfo {author} {\bibfnamefont {Q.}~\bibnamefont {Liang}},
  \bibinfo {author} {\bibfnamefont {B.}~\bibnamefont {Zhang}}, \bibinfo
  {author} {\bibfnamefont {S.}~\bibnamefont {Huang}}, \bibinfo {author}
  {\bibfnamefont {Y.}~\bibnamefont {Wang}}, \bibinfo {author} {\bibfnamefont
  {C.}~\bibnamefont {Fang}}, \bibinfo {author} {\bibfnamefont {J.}~\bibnamefont
  {Kim}},\ and\ \bibinfo {author} {\bibfnamefont {K.~R.}\ \bibnamefont
  {Brown}},\ }\bibfield  {title} {\bibinfo {title} {Batch optimization of
  frequency-modulated pulses for robust two-qubit gates in ion chains},\ }\href
  {https://doi.org/10.1103/PhysRevApplied.16.024039} {\bibfield  {journal}
  {\bibinfo  {journal} {Phys. Rev. Applied}\ }\textbf {\bibinfo {volume}
  {16}},\ \bibinfo {pages} {024039} (\bibinfo {year} {2021})}\BibitemShut
  {NoStop}%
\bibitem [{Note1()}]{Note1}%
  \BibitemOpen
  \bibinfo {note} {Unlike the coherent crosstalk noise caused by a single
  addressing beam, the finite phase coherence time between two addressing beams
  makes the interbeam crosstalk stochastic noise after averaging over multiple
  experiment shots.}\BibitemShut {Stop}%
\bibitem [{\citenamefont {Jones}(2003)}]{JonesPRA2003}%
  \BibitemOpen
  \bibfield  {author} {\bibinfo {author} {\bibfnamefont {J.~A.}\ \bibnamefont
  {Jones}},\ }\bibfield  {title} {\bibinfo {title} {Robust ising gates for
  practical quantum computation},\ }\href
  {https://doi.org/10.1103/PhysRevA.67.012317} {\bibfield  {journal} {\bibinfo
  {journal} {Phys. Rev. A}\ }\textbf {\bibinfo {volume} {67}},\ \bibinfo
  {pages} {012317} (\bibinfo {year} {2003})}\BibitemShut {NoStop}%
\bibitem [{\citenamefont {Tomita}\ \emph {et~al.}(2010)\citenamefont {Tomita},
  \citenamefont {Merrill},\ and\ \citenamefont {Brown}}]{TomitaNJP2010}%
  \BibitemOpen
  \bibfield  {author} {\bibinfo {author} {\bibfnamefont {Y.}~\bibnamefont
  {Tomita}}, \bibinfo {author} {\bibfnamefont {J.~T.}\ \bibnamefont
  {Merrill}},\ and\ \bibinfo {author} {\bibfnamefont {K.~R.}\ \bibnamefont
  {Brown}},\ }\bibfield  {title} {\bibinfo {title} {Multi-qubit compensation
  sequences},\ }\href {https://doi.org/10.1088/1367-2630/12/1/015002}
  {\bibfield  {journal} {\bibinfo  {journal} {New Journal of Physics}\ }\textbf
  {\bibinfo {volume} {12}},\ \bibinfo {pages} {015002} (\bibinfo {year}
  {2010})}\BibitemShut {NoStop}%
\bibitem [{\citenamefont {Murphy}\ and\ \citenamefont
  {Brown}(2019)}]{murphy2019}%
  \BibitemOpen
  \bibfield  {author} {\bibinfo {author} {\bibfnamefont {D.~C.}\ \bibnamefont
  {Murphy}}\ and\ \bibinfo {author} {\bibfnamefont {K.~R.}\ \bibnamefont
  {Brown}},\ }\bibfield  {title} {\bibinfo {title} {Controlling error
  orientation to improve quantum algorithm success rates},\ }\href
  {https://doi.org/10.1103/PhysRevA.99.032318} {\bibfield  {journal} {\bibinfo
  {journal} {Phys. Rev. A}\ }\textbf {\bibinfo {volume} {99}},\ \bibinfo
  {pages} {032318} (\bibinfo {year} {2019})}\BibitemShut {NoStop}%
\bibitem [{\citenamefont {Brown}\ \emph {et~al.}(2006)\citenamefont {Brown},
  \citenamefont {Clark},\ and\ \citenamefont {Chuang}}]{BrownPRL2006}%
  \BibitemOpen
  \bibfield  {author} {\bibinfo {author} {\bibfnamefont {K.~R.}\ \bibnamefont
  {Brown}}, \bibinfo {author} {\bibfnamefont {R.~J.}\ \bibnamefont {Clark}},\
  and\ \bibinfo {author} {\bibfnamefont {I.~L.}\ \bibnamefont {Chuang}},\
  }\bibfield  {title} {\bibinfo {title} {Limitations of quantum simulation
  examined by simulating a pairing hamiltonian using nuclear magnetic
  resonance},\ }\href {https://doi.org/10.1103/PhysRevLett.97.050504}
  {\bibfield  {journal} {\bibinfo  {journal} {Phys. Rev. Lett.}\ }\textbf
  {\bibinfo {volume} {97}},\ \bibinfo {pages} {050504} (\bibinfo {year}
  {2006})}\BibitemShut {NoStop}%
\bibitem [{\citenamefont {Nam}\ \emph {et~al.}(2020)\citenamefont {Nam},
  \citenamefont {Chen}, \citenamefont {Pisenti}, \citenamefont {Wright},
  \citenamefont {Delaney}, \citenamefont {Maslov}, \citenamefont {Brown},
  \citenamefont {Allen}, \citenamefont {Amini}, \citenamefont {Apisdorf} \emph
  {et~al.}}]{NamNQuantumInf2019groundstate}%
  \BibitemOpen
  \bibfield  {author} {\bibinfo {author} {\bibfnamefont {Y.}~\bibnamefont
  {Nam}}, \bibinfo {author} {\bibfnamefont {J.-S.}\ \bibnamefont {Chen}},
  \bibinfo {author} {\bibfnamefont {N.~C.}\ \bibnamefont {Pisenti}}, \bibinfo
  {author} {\bibfnamefont {K.}~\bibnamefont {Wright}}, \bibinfo {author}
  {\bibfnamefont {C.}~\bibnamefont {Delaney}}, \bibinfo {author} {\bibfnamefont
  {D.}~\bibnamefont {Maslov}}, \bibinfo {author} {\bibfnamefont {K.~R.}\
  \bibnamefont {Brown}}, \bibinfo {author} {\bibfnamefont {S.}~\bibnamefont
  {Allen}}, \bibinfo {author} {\bibfnamefont {J.~M.}\ \bibnamefont {Amini}},
  \bibinfo {author} {\bibfnamefont {J.}~\bibnamefont {Apisdorf}}, \emph
  {et~al.},\ }\bibfield  {title} {\bibinfo {title} {Ground-state energy
  estimation of the water molecule on a trapped-ion quantum computer},\
  }\href@noop {} {\bibfield  {journal} {\bibinfo  {journal} {npj Quantum
  Information}\ }\textbf {\bibinfo {volume} {6}},\ \bibinfo {pages} {1}
  (\bibinfo {year} {2020})}\BibitemShut {NoStop}%
\bibitem [{\citenamefont {O'Malley}\ \emph {et~al.}(2016)\citenamefont
  {O'Malley}, \citenamefont {Babbush}, \citenamefont {Kivlichan}, \citenamefont
  {Romero}, \citenamefont {McClean}, \citenamefont {Barends}, \citenamefont
  {Kelly}, \citenamefont {Roushan}, \citenamefont {Tranter}, \citenamefont
  {Ding}, \citenamefont {Campbell}, \citenamefont {Chen}, \citenamefont {Chen},
  \citenamefont {Chiaro}, \citenamefont {Dunsworth}, \citenamefont {Fowler},
  \citenamefont {Jeffrey}, \citenamefont {Lucero}, \citenamefont {Megrant},
  \citenamefont {Mutus}, \citenamefont {Neeley}, \citenamefont {Neill},
  \citenamefont {Quintana}, \citenamefont {Sank}, \citenamefont {Vainsencher},
  \citenamefont {Wenner}, \citenamefont {White}, \citenamefont {Coveney},
  \citenamefont {Love}, \citenamefont {Neven}, \citenamefont {Aspuru-Guzik},\
  and\ \citenamefont {Martinis}}]{OMalleyPRX2016}%
  \BibitemOpen
  \bibfield  {author} {\bibinfo {author} {\bibfnamefont {P.~J.~J.}\
  \bibnamefont {O'Malley}}, \bibinfo {author} {\bibfnamefont {R.}~\bibnamefont
  {Babbush}}, \bibinfo {author} {\bibfnamefont {I.~D.}\ \bibnamefont
  {Kivlichan}}, \bibinfo {author} {\bibfnamefont {J.}~\bibnamefont {Romero}},
  \bibinfo {author} {\bibfnamefont {J.~R.}\ \bibnamefont {McClean}}, \bibinfo
  {author} {\bibfnamefont {R.}~\bibnamefont {Barends}}, \bibinfo {author}
  {\bibfnamefont {J.}~\bibnamefont {Kelly}}, \bibinfo {author} {\bibfnamefont
  {P.}~\bibnamefont {Roushan}}, \bibinfo {author} {\bibfnamefont
  {A.}~\bibnamefont {Tranter}}, \bibinfo {author} {\bibfnamefont
  {N.}~\bibnamefont {Ding}}, \bibinfo {author} {\bibfnamefont {B.}~\bibnamefont
  {Campbell}}, \bibinfo {author} {\bibfnamefont {Y.}~\bibnamefont {Chen}},
  \bibinfo {author} {\bibfnamefont {Z.}~\bibnamefont {Chen}}, \bibinfo {author}
  {\bibfnamefont {B.}~\bibnamefont {Chiaro}}, \bibinfo {author} {\bibfnamefont
  {A.}~\bibnamefont {Dunsworth}}, \bibinfo {author} {\bibfnamefont {A.~G.}\
  \bibnamefont {Fowler}}, \bibinfo {author} {\bibfnamefont {E.}~\bibnamefont
  {Jeffrey}}, \bibinfo {author} {\bibfnamefont {E.}~\bibnamefont {Lucero}},
  \bibinfo {author} {\bibfnamefont {A.}~\bibnamefont {Megrant}}, \bibinfo
  {author} {\bibfnamefont {J.~Y.}\ \bibnamefont {Mutus}}, \bibinfo {author}
  {\bibfnamefont {M.}~\bibnamefont {Neeley}}, \bibinfo {author} {\bibfnamefont
  {C.}~\bibnamefont {Neill}}, \bibinfo {author} {\bibfnamefont
  {C.}~\bibnamefont {Quintana}}, \bibinfo {author} {\bibfnamefont
  {D.}~\bibnamefont {Sank}}, \bibinfo {author} {\bibfnamefont {A.}~\bibnamefont
  {Vainsencher}}, \bibinfo {author} {\bibfnamefont {J.}~\bibnamefont {Wenner}},
  \bibinfo {author} {\bibfnamefont {T.~C.}\ \bibnamefont {White}}, \bibinfo
  {author} {\bibfnamefont {P.~V.}\ \bibnamefont {Coveney}}, \bibinfo {author}
  {\bibfnamefont {P.~J.}\ \bibnamefont {Love}}, \bibinfo {author}
  {\bibfnamefont {H.}~\bibnamefont {Neven}}, \bibinfo {author} {\bibfnamefont
  {A.}~\bibnamefont {Aspuru-Guzik}},\ and\ \bibinfo {author} {\bibfnamefont
  {J.~M.}\ \bibnamefont {Martinis}},\ }\bibfield  {title} {\bibinfo {title}
  {Scalable quantum simulation of molecular energies},\ }\href
  {https://doi.org/10.1103/PhysRevX.6.031007} {\bibfield  {journal} {\bibinfo
  {journal} {Phys. Rev. X}\ }\textbf {\bibinfo {volume} {6}},\ \bibinfo {pages}
  {031007} (\bibinfo {year} {2016})}\BibitemShut {NoStop}%
\bibitem [{\citenamefont {Martinez}\ \emph {et~al.}(2016)\citenamefont
  {Martinez}, \citenamefont {Monz}, \citenamefont {Nigg}, \citenamefont
  {Schindler},\ and\ \citenamefont {Blatt}}]{martinez2016compiling}%
  \BibitemOpen
  \bibfield  {author} {\bibinfo {author} {\bibfnamefont {E.~A.}\ \bibnamefont
  {Martinez}}, \bibinfo {author} {\bibfnamefont {T.}~\bibnamefont {Monz}},
  \bibinfo {author} {\bibfnamefont {D.}~\bibnamefont {Nigg}}, \bibinfo {author}
  {\bibfnamefont {P.}~\bibnamefont {Schindler}},\ and\ \bibinfo {author}
  {\bibfnamefont {R.}~\bibnamefont {Blatt}},\ }\bibfield  {title} {\bibinfo
  {title} {Compiling quantum algorithms for architectures with multi-qubit
  gates},\ }\href@noop {} {\bibfield  {journal} {\bibinfo  {journal} {New
  Journal of Physics}\ }\textbf {\bibinfo {volume} {18}},\ \bibinfo {pages}
  {063029} (\bibinfo {year} {2016})}\BibitemShut {NoStop}%
\bibitem [{\citenamefont {Cincio}\ \emph {et~al.}(2021)\citenamefont {Cincio},
  \citenamefont {Rudinger}, \citenamefont {Sarovar},\ and\ \citenamefont
  {Coles}}]{CincioPRXQuantum2021}%
  \BibitemOpen
  \bibfield  {author} {\bibinfo {author} {\bibfnamefont {L.}~\bibnamefont
  {Cincio}}, \bibinfo {author} {\bibfnamefont {K.}~\bibnamefont {Rudinger}},
  \bibinfo {author} {\bibfnamefont {M.}~\bibnamefont {Sarovar}},\ and\ \bibinfo
  {author} {\bibfnamefont {P.~J.}\ \bibnamefont {Coles}},\ }\bibfield  {title}
  {\bibinfo {title} {Machine learning of noise-resilient quantum circuits},\
  }\href {https://doi.org/10.1103/PRXQuantum.2.010324} {\bibfield  {journal}
  {\bibinfo  {journal} {PRX Quantum}\ }\textbf {\bibinfo {volume} {2}},\
  \bibinfo {pages} {010324} (\bibinfo {year} {2021})}\BibitemShut {NoStop}%
\bibitem [{\citenamefont {Sharma}\ \emph {et~al.}(2020)\citenamefont {Sharma},
  \citenamefont {Khatri}, \citenamefont {Cerezo},\ and\ \citenamefont
  {Coles}}]{SharmaNJP2020}%
  \BibitemOpen
  \bibfield  {author} {\bibinfo {author} {\bibfnamefont {K.}~\bibnamefont
  {Sharma}}, \bibinfo {author} {\bibfnamefont {S.}~\bibnamefont {Khatri}},
  \bibinfo {author} {\bibfnamefont {M.}~\bibnamefont {Cerezo}},\ and\ \bibinfo
  {author} {\bibfnamefont {P.~J.}\ \bibnamefont {Coles}},\ }\bibfield  {title}
  {\bibinfo {title} {Noise resilience of variational quantum compiling},\
  }\href@noop {} {\bibfield  {journal} {\bibinfo  {journal} {New Journal of
  Physics}\ }\textbf {\bibinfo {volume} {22}},\ \bibinfo {pages} {043006}
  (\bibinfo {year} {2020})}\BibitemShut {NoStop}%
\bibitem [{\citenamefont {Yeter-Aydeniz}\ \emph {et~al.}(2021)\citenamefont
  {Yeter-Aydeniz}, \citenamefont {Gard}, \citenamefont {Jakowski},
  \citenamefont {Majumder}, \citenamefont {Barron}, \citenamefont {Siopsis},
  \citenamefont {Humble},\ and\ \citenamefont
  {Pooser}}]{yeter2021benchmarking}%
  \BibitemOpen
  \bibfield  {author} {\bibinfo {author} {\bibfnamefont {K.}~\bibnamefont
  {Yeter-Aydeniz}}, \bibinfo {author} {\bibfnamefont {B.~T.}\ \bibnamefont
  {Gard}}, \bibinfo {author} {\bibfnamefont {J.}~\bibnamefont {Jakowski}},
  \bibinfo {author} {\bibfnamefont {S.}~\bibnamefont {Majumder}}, \bibinfo
  {author} {\bibfnamefont {G.~S.}\ \bibnamefont {Barron}}, \bibinfo {author}
  {\bibfnamefont {G.}~\bibnamefont {Siopsis}}, \bibinfo {author} {\bibfnamefont
  {T.~S.}\ \bibnamefont {Humble}},\ and\ \bibinfo {author} {\bibfnamefont
  {R.~C.}\ \bibnamefont {Pooser}},\ }\bibfield  {title} {\bibinfo {title}
  {Benchmarking quantum chemistry computations with variational, imaginary time
  evolution, and krylov space solver algorithms},\ }\href@noop {} {\bibfield
  {journal} {\bibinfo  {journal} {Advanced Quantum Technologies}\ }\textbf
  {\bibinfo {volume} {4}},\ \bibinfo {pages} {2100012} (\bibinfo {year}
  {2021})}\BibitemShut {NoStop}%
\bibitem [{\citenamefont {Hayes}(2012)}]{hayes2012remote}%
  \BibitemOpen
  \bibfield  {author} {\bibinfo {author} {\bibfnamefont {D.~L.}\ \bibnamefont
  {Hayes}},\ }\href@noop {} {\emph {\bibinfo {title} {Remote and local
  entanglement of ions using photons and phonons}}}\ (\bibinfo  {publisher}
  {University of Maryland, College Park},\ \bibinfo {year} {2012})\BibitemShut
  {NoStop}%
\bibitem [{\citenamefont {Gardiner}\ \emph {et~al.}(2004)\citenamefont
  {Gardiner}, \citenamefont {Zoller},\ and\ \citenamefont
  {Zoller}}]{gardiner2004quantum}%
  \BibitemOpen
  \bibfield  {author} {\bibinfo {author} {\bibfnamefont {C.}~\bibnamefont
  {Gardiner}}, \bibinfo {author} {\bibfnamefont {P.}~\bibnamefont {Zoller}},\
  and\ \bibinfo {author} {\bibfnamefont {P.}~\bibnamefont {Zoller}},\
  }\href@noop {} {\emph {\bibinfo {title} {Quantum noise: a handbook of
  Markovian and non-Markovian quantum stochastic methods with applications to
  quantum optics}}}\ (\bibinfo  {publisher} {Springer Science \& Business
  Media},\ \bibinfo {year} {2004})\BibitemShut {NoStop}%
\bibitem [{\citenamefont {Lindblad}(1976)}]{lindblad1976generators}%
  \BibitemOpen
  \bibfield  {author} {\bibinfo {author} {\bibfnamefont {G.}~\bibnamefont
  {Lindblad}},\ }\bibfield  {title} {\bibinfo {title} {On the generators of
  quantum dynamical semigroups},\ }\href@noop {} {\bibfield  {journal}
  {\bibinfo  {journal} {Communications in Mathematical Physics}\ }\textbf
  {\bibinfo {volume} {48}},\ \bibinfo {pages} {119} (\bibinfo {year}
  {1976})}\BibitemShut {NoStop}%
\bibitem [{\citenamefont {Johansson}\ \emph {et~al.}(2012)\citenamefont
  {Johansson}, \citenamefont {Nation},\ and\ \citenamefont
  {Nori}}]{johansson2012qutip}%
  \BibitemOpen
  \bibfield  {author} {\bibinfo {author} {\bibfnamefont {J.~R.}\ \bibnamefont
  {Johansson}}, \bibinfo {author} {\bibfnamefont {P.~D.}\ \bibnamefont
  {Nation}},\ and\ \bibinfo {author} {\bibfnamefont {F.}~\bibnamefont {Nori}},\
  }\bibfield  {title} {\bibinfo {title} {Qutip: An open-source python framework
  for the dynamics of open quantum systems},\ }\href@noop {} {\bibfield
  {journal} {\bibinfo  {journal} {Comput. Phys. Commun.}\ }\textbf {\bibinfo
  {volume} {183}},\ \bibinfo {pages} {1760} (\bibinfo {year}
  {2012})}\BibitemShut {NoStop}%
\bibitem [{\citenamefont {Greenbaum}(2015)}]{greenbaum2015introduction}%
  \BibitemOpen
  \bibfield  {author} {\bibinfo {author} {\bibfnamefont {D.}~\bibnamefont
  {Greenbaum}},\ }\bibfield  {title} {\bibinfo {title} {Introduction to quantum
  gate set tomography},\ }\href@noop {} {\bibfield  {journal} {\bibinfo
  {journal} {arXiv preprint arXiv:1509.02921}\ } (\bibinfo {year}
  {2015})}\BibitemShut {NoStop}%
\bibitem [{\citenamefont {Nielsen}(1996)}]{Nielsen1996}%
  \BibitemOpen
  \bibfield  {author} {\bibinfo {author} {\bibfnamefont {M.~A.}\ \bibnamefont
  {Nielsen}},\ }\bibfield  {title} {\bibinfo {title} {The entanglement fidelity
  and quantum error correction},\ }\href@noop {} {\bibfield  {journal}
  {\bibinfo  {journal} {arXiv preprint arXiv:quant-ph/9606012}\ } (\bibinfo
  {year} {1996})}\BibitemShut {NoStop}%
\end{thebibliography}%
\end{document}